\documentclass[10pt,twocolumn]{IEEEtran}
\usepackage{cite}

\usepackage{amsmath,amssymb,amsfonts}
\usepackage[linesnumbered,ruled,vlined]{algorithm2e}
\usepackage[dvips]{graphicx}
\usepackage{textcomp}
\usepackage{xcolor}\usepackage[T1]{fontenc}
\usepackage{colortbl}
\usepackage{titlesec}
\usepackage{soul}
\usepackage{subfigure}
\usepackage{url}  
\usepackage[bottom]{footmisc}
\usepackage{verbatim,float,caption}
\AtBeginDocument{\AtBeginShipoutNext{\AtBeginShipoutDiscard}} 
\usepackage{hyperref} 
\usepackage{flushend}  
\begin{document}
	
	\title{Dynamic Fairness-Aware Spectrum Auction for Enhanced Licensed Shared Access in 6G Networks}
	\author{\IEEEauthorblockN{Mina Khadem, Maryam Ansarifard, Nader Mokari, \IEEEmembership{Senior Member, IEEE}, Mohammadreza Javan, \IEEEmembership{Senior Member, IEEE}, Hamid Saeedi, \IEEEmembership{Senior Member, IEEE}
	}, Eduard A. Jorswieck, \IEEEmembership{Fellow, IEEE}}
	\thanks{
	{M. Khadem, M. Ansarifard, N. Mokari and H. Saeedi are with the Department of Electrical and Computer Engineering, Tarbiat Modares University, Tehran, Iran, (emails: {mina.khadem, m.ansarifard, nader.mokari, hsaeedi}@modares.ac.ir). H. Saeedi is also with University of Doha for Science and Technology, Doha, Qatar. M. R. Javan is with the Department of Electrical and Computer Engineering, Shahrood University of Technology, Iran, (email: javan@shahroodut.ac.ir). Eduard A. Jorswieck is with Institute for Communications Technology, TU Braunschweig, Germany,
	    email: jorswieck@ifn.ing.tu-bs.de.}}
	\maketitle
	\markboth{IEEE Transactions on Communications}%
    {Submitted paper}
	\begin{abstract} 
This article introduces a new approach to address the spectrum scarcity challenge in 6G networks by implementing the enhanced licensed shared access (ELSA) framework. \textcolor{black}{Our} proposed auction mechanism aims to ensure fairness in spectrum allocation to mobile network operators (MNOs) through a novel weighted auction called the fair Vickery-Clarke-Groves (FVCG) mechanism. Through comparison with traditional methods, the study demonstrates that the proposed auction method improves fairness significantly. We suggest using spectrum sensing and integrating UAV-based networks to enhance efficiency of the LSA system. This research employs two methods to solve the problem. We first propose a novel greedy algorithm, named market share-based weighted greedy algorithm (MSWGA) to achieve better fairness compared to the traditional auction methods and as the second approach, we exploit deep reinforcement learning (DRL) algorithms, to optimize the auction policy and demonstrate its superiority over other methods.
Simulation results show that the deep deterministic policy gradient (DDPG) method performs superior to soft actor critic (SAC), MSWGA, and greedy methods. Moreover, a significant improvement is observed in fairness index compared to the traditional greedy auction methods. 
This improvement is as high as 
about 27\% and 35\% when deploying  the MSWGA and DDPG methods, respectively.  
	\end{abstract}
	\section{INTRODUCTION}
	\subsection{Motivation}
	\IEEEPARstart{T}{he} next generation of wireless communication systems comprises many terrestrial, aerial, and space/satellite platforms that function independently and are called three dimensional (3D) networks. Its functionality is more dynamic than today's fixed terrestrial cellular network. Due to the increasing number of users and the crucial role of wireless networks in the future, numerous capabilities of unmanned aerial vehicles (UAVs), including mobility and the creation of line of sight (LoS) links, have made it an applicable option for 5G and beyond \cite{zhao2020efficiency}. Its inherent characteristics allow to have a feasible network that can play a \textcolor{black}{favorable} role in emergencies, natural disasters, and high-traffic areas. Thus, the role of UAV-based 3D networks in next generation networks and their impact on reducing network traffic is essential \cite{geraci2022integrating}.
	
	\textcolor{black}{Considering the rapid growth of wireless networks and future spectrum scarcity, spectrum sharing emerges as a suitable solution for effectively utilizing this non-renewable and valuable resource}. One of the methods for spectrum sharing is licensed shared access (LSA) \cite{ecc205}, which aims to alleviate network traffic and add the existing spectrum of mobile network operators (MNOs) by utilizing licensed bands for a specified duration. Sharing spectrum on a larger scale, like operators, could be regarded as a potential gain in network efficiency.
    However, the LSA method faces challenges, such as accurately identifying vacant spectrum in real-time, which may reduce spectral efficiency (SE) \cite{matinmikko2014spectrum}.While the LSA method allows MNOs to share allocated spectrum with predefined rules, the conventional static exclusive licensing approach, governed by  national regulatory authorities (NRAs), may result in inefficient spectrum utilization. Therefore, the collaborative research project ADEL, funded by the European Union, suggests that in addition to supporting the traditional LSA network architecture, it uses sensing to improve spectrum utilization (see Fig.1) \cite{frascolla2016dynamic}. Such an enhanced LSA network is referred to, in this paper, as \textcolor{black}{enhanced LSA (ELSA).}
    
    Based on the discussions surrounding the utilization of 3D networks in upcoming mobile network generations, as well as the remarkable application of UAVs within these networks, it is favorable to consider their deployment as spectrum sensing components within the suggested ADEL network \cite{morgado2015dynamic}.
    On the other hand, an auction-based dynamic spectrum allocation method can lead to a fair approach for allocating spectrum among diverse users \cite{niyato2020auction}. Therefore, an auction-based spectrum allocation method, ensuring fairness, became a suitable approach to enhance spectrum efficiency and reduce MNO traffic as needed. Unlike a bundled auction where all spectrum blocks are sold together at a fixed price, employing a dynamic auction can influence the strategies of other buyers in the spectrum.

	\subsection{Related Works}
	
		\begin{table*}[t]
		\centering
		\caption{Summary of the related works}
		\label{summary of related works}
		\scalebox{0.77}{
			\begin{tabular}{|c|c|c|c|c|c|c|}
				\rowcolor{blue!20}
				\hline
				\textbf{Ref.} & \textbf{\begin{tabular}[c]{@{}c@{}}Objective Function\end{tabular}} & \textbf{\begin{tabular}[c]{@{}c@{}}Optimization Algorithm\end{tabular}} & \textbf{Auction} & \textbf{Auction Mechanism} & \textbf{\begin{tabular}[c]{@{}c@{}}Auction\\ Fairness\end{tabular}} & \textbf{\begin{tabular}[c]{@{}c@{}}Spectrum\\ Sensing\end{tabular}} \\ \hline
				\cite{shang2020spectrum} & Maximizing spectrum efficiency & Blockchain-based & $\color{red}{\pmb{\mathsf{x}}}$ & $\color{red}{\pmb{\mathsf{x}}}$ & $\color{red}{\pmb{\mathsf{x}}}$ & $\color{green}{\pmb{\mathsf{\checkmark}}}$ \\ \hline
				\cite{onidare2020spectral} & Maximizing spectrum efficiency & Heuristic & $\color{red}{\pmb{\mathsf{x}}}$ & $\color{red}{\pmb{\mathsf{x}}}$ & $\color{red}{\pmb{\mathsf{x}}}$ & $\color{red}{\pmb{\mathsf{x}}}$ \\ \hline
				\cite{butt2018fair} & Maximizing allocated spectrum & Heuristic & $\color{red}{\pmb{\mathsf{x}}}$ & $\color{red}{\pmb{\mathsf{x}}}$ & $\color{green}{\pmb{\mathsf{\checkmark}}}$ & $\color{red}{\pmb{\mathsf{x}}}$ \\ \hline
				\cite{kokkinen2019results} & Minimizing interference & Correlation algorithm & $\color{red}{\pmb{\mathsf{x}}}$ & $\color{red}{\pmb{\mathsf{x}}}$ & $\color{red}{\pmb{\mathsf{x}}}$ & $\color{green}{\pmb{\mathsf{\checkmark}}}$ \\ \hline
				\cite{shang20203d} & Maximizing spectrum efficiency & Monte-carlo & $\color{red}{\pmb{\mathsf{x}}}$ & $\color{red}{\pmb{\mathsf{x}}}$ & $\color{red}{\pmb{\mathsf{x}}}$ & $\color{green}{\pmb{\mathsf{\checkmark}}}$ \\ \hline
				\cite{liu2018spectrum} & Maximizing operational power & Alternating direction optimization (ADO) & $\color{red}{\pmb{\mathsf{x}}}$ & $\color{red}{\pmb{\mathsf{x}}}$ & $\color{red}{\pmb{\mathsf{x}}}$ & $\color{green}{\pmb{\mathsf{\checkmark}}}$ \\ \hline
				\cite{wu2021optimisation} & Maximizing operational power & Numerical simulation & $\color{red}{\pmb{\mathsf{x}}}$ & $\color{red}{\pmb{\mathsf{x}}}$ & $\color{red}{\pmb{\mathsf{x}}}$ & $\color{green}{\pmb{\mathsf{\checkmark}}}$ \\ \hline
				\cite{chouayakh2020designing} & Maximizing social welfare (income and fairness) & Greedy & $\color{green}{\pmb{\mathsf{\checkmark}}}$ & VCG mechanism & $\color{green}{\pmb{\mathsf{\checkmark}}}$ & $\color{red}{\pmb{\mathsf{x}}}$ \\ \hline
				\cite{devi2021double} & Maximizing allocated spectrum & Greedy & $\color{green}{\pmb{\mathsf{\checkmark}}}$ & Sealed-bid double auction & $\color{red}{\pmb{\mathsf{x}}}$ & $\color{green}{\pmb{\mathsf{\checkmark}}}$ \\ \hline
				\cite{chouayakh2021ascending} & Maximizing allocated resources (spectrum and antennas) & Greedy & $\color{green}{\pmb{\mathsf{\checkmark}}}$ & Clock auction & $\color{red}{\pmb{\mathsf{x}}}$ & $\color{red}{\pmb{\mathsf{x}}}$ \\ \hline
				\cite{sujit2007distributed} & Minimizing costs & Greedy & $\color{green}{\pmb{\mathsf{\checkmark}}}$ & Sequential auction & $\color{red}{\pmb{\mathsf{x}}}$ & $\color{green}{\pmb{\mathsf{\checkmark}}}$ \\ \hline
				\cite{feng2014flexauc} & Maximizing revenue & Greedy & $\color{green}{\pmb{\mathsf{\checkmark}}}$ & Heuristic (flex auction) & $\color{red}{\pmb{\mathsf{x}}}$ & $\color{red}{\pmb{\mathsf{x}}}$ \\ \hline
				\cite{csercsik2023preallocation} & Maximizing the total utility of the system & Greedy & $\color{green}{\pmb{\mathsf{\checkmark}}}$ & Preallocation-based combinatorial auction & $\color{green}{\pmb{\mathsf{\checkmark}}}$ & $\color{red}{\pmb{\mathsf{x}}}$ \\ \hline
				Our Work & Maximizing fairness index & RL-based (DDPG) & $\color{green}{\pmb{\mathsf{\checkmark}}}$ & Fair Vickrey–Clarke–Groves (FVCG) & $\color{green}{\pmb{\mathsf{\checkmark}}}$ & $\color{green}{\pmb{\mathsf{\checkmark}}}$ \\ \hline
			\end{tabular}
		}
	\end{table*}
	The existing literature can be categorized into five separate sections, considering the comprehensive nature of the content and the diverse range of topics covered in this paper:
	1) LSA-based spectrum sharing; 2) Auction-based markets for resource allocation; 3) UAV-based integrated networks; 4) Spectrum sensing using UAVs; 5) RL-based cellular networks simulation.\\
	
	\emph{\indent \textbf{1) LSA-based spectrum sharing:}}
	LSA is a suitable candidate to address spectrum scarcity in populated areas; for instance, the authors in \cite{onidare2020spectral} use an optimization problem to determine the optimal uplink and downlink power allocation in the licensee. The objective is to maximize the licensee's SE subject to the incumbent interference thresholds. In \cite{butt2018fair}, a novel spectrum management algorithm is introduced to facilitate dynamic spectrum sharing between a primary main network and \textcolor{black}{
	MNOs as secondary users.} The authors put forward a mechanism that takes into account regulatory compliance and imposes penalties on MNOs who have neglected their responsibility to allocate spectrum appropriately. Ref. 
	\cite{kokkinen2019results} examines the difficulties associated with ensuring an adequate allocation of spectrum for various users. It introduces LSA as a technological solution for effective spectrum sharing. Moreover, the paper proposes the utilization of spectrum sensing techniques to mitigate potential interference.
	In \cite{chouayakh2020designing}, the authors explore the utilization of shared licenses as a viable solution for MNOs to gain access to the existing spectrum through suitable licensing procedures. They emphasize the significance of allocating the spectrum efficiently through auctions, taking into account considerations related to radio interference. Furthermore, the paper undertakes a comprehensive examination of the existing mechanisms, puts forward novel mechanisms, and assesses advancements in interference management methods for base stations.
	\\	
	\emph{\indent \textbf{2) Auction-based markets for resource allocation:}} 
	Auctions are efficient mechanisms for allocating goods, tasks, and infrastructures in time-limited situations. For instance, in \cite{chouayakh2018pam}, a truthful auction mechanism called the proportional allocation mechanism (PAM) provided a framework that is able to balance the compromise between fairness and efficiency. In\cite{butt2018fair}, a penalty-based fair spectrum allocation algorithm for distributing the available spectrum of incumbent among MNOs is proposed. The author in \cite{devi2021double} use a sealed-bid double auction to improve network performance in  cognitive
	radio (CR) technology and network dynamics on the secondary user side. In \cite{mcmenamy2016enhanced}, an approach is employed to derive a function of the minimum average required rate and the antenna price, where the spectrum can be leased in blocks by the virtual network operator (VNO). The study proposes an auction-based LSA framework that enables VNOs to share spectrum and infrastructure, including antennas, using technologies like massive multiple-input multiple-output (MIMO) and virtualization to enhance the framework \textcolor{black}{while ensuring quality of service (QoS) for all users}. Paper \cite{chouayakh2021ascending} explores LSA, allowing MNOs to share bandwidth in the $2.3-2.4$ GHz spectrum range with the incumbent. The authors propose an ascending auction mechanism based on Vickrey-Clarke-Groves (VCG) auction to allocate spectrum among base station groups, ensuring fairness and truthfulness. The proposed mechanism's efficiency is evaluated in terms of social welfare.
	In \cite{feng2014flexauc}, auctions in secondary spectrum trading are discussed, where spectrum holders (SH) redistribute unused channels to wireless service providers (WSP). The paper examines strategies of WSPs, SHs, and end users in maximizing revenue and providing effective services.
	\\
	\emph{\indent \textbf{3) UAV-based integrated networks:}}
		Authors in \cite{shang20203d} focus on the exploration of the 3D spectrum sharing between device-to-device communication (D2D) and UAVs. The authors investigate the utilization of UAVs for spatial spectrum sensing, enabling opportunistic access to channels occupied by D2D communications of ground users. The primary objective is to optimize the 3D spectrum sharing, ensuring maximum area spectral efficiency (ASE) for UAV networks while maintaining the minimum required ASE for D2D networks.
    	Ref. \cite{santana2018cognitive} provides a comprehensive review of CR technology for communication with UAVs in unlicensed frequency bands. As UAVs are increasingly employed in various applications, including the internet of things (IoT), the existing unlicensed bands become crowded. To address this issue, the paper suggests leveraging CR techniques and spectrum sharing as promising solutions to alleviate spectrum scarcity in wireless networks, thereby supporting the development of future wireless networks.
    In \cite{cheng2016auction}, the authors address the task allocation problem within a multi-UAV system, taking into account various limitations and constraints. To do so, the authors propose a method based on the auction algorithm, wherein UAVs compete for tasks by submitting proposals. This approach incorporates a multi-layer cost calculation method, dividing the cost into four layers associated with different types of restrictions.
    In \cite{liu2021spectrum}, the authors highlight sixth-generation (6G) networks and high-power satellite communication services as potential solutions to achieve comprehensive and high-speed network access. The integrated satellite-terrestrial communication networks (ISTCN) are introduced, aiming to integrate terrestrial networks with satellite communication networks to provide seamless and integrated network services. The paper introduces techniques to enhance spectrum sharing in ISTCN, employing non-orthogonal multiple access (NOMA) and CR technology.
    Lastly, \cite{wang2020robust} focuses on the potential of integrated air-ground networks (AGIN) and the limitations posed by limited spectrum resources. Spectrum sharing is considered as a technique to improve spectrum utilization in AGIN, presenting new opportunities and challenges.
	\\
	\emph{\indent \textbf{4) Spectrum sensing using UAVs:}}
	In \cite{liu2018spectrum}, the authors propose a CR system model that utilizes one UAV. The aim is to enhance spectrum sensing performance and access to idle spectrum. The system model operates by deploying the UAV in a circular path around the primary user (PU), dividing the flight period into sensing and transmission phases. During the sensing phase, the UAV can access the spectrum only if no PU is present. The authors present simulation results demonstrating the improved transmission performance of the UAV through optimal spectrum sensing and superiority in spectrum sensing.
	In a similar way, the authors in \cite{wu2021optimisation} investigate a CR system based on a UAV, where the UAV functions as a secondary user. It utilizes spectrum measurement capabilities to transmit data.
	Furthermore, \cite{zhang2009optimization} focuses on cooperative spectrum sensing, wherein multiple CRs collaborate to identify available spectrum blocks using energy detection. The objective is to optimize the diagnosis performance in an efficient and practical manner by examining the optimal participation spectrum sensing.
	\\
	\emph{\indent\textbf{5) RL-based cellular networks simulation:}} 
	In \cite{niyato2008competitive}, the authors propose a spectral pricing method based on game theory, where the final price is determined through Nash equilibrium.
	In \cite{dutting2019optimal}, the authors explore the application of machine learning (ML) tools to automate the design of optimal auctions. They model the auction as a multi-layer neural network, focusing on settings involving multiple bids and items.
	\cite{shamsoshoara2019distributed} presents a distributed mechanism for spectrum sharing between UAVs and licensed terrestrial networks. The study considers a centralized scenario for remote sensing missions, where UAVs are divided into two clusters: relay UAVs and sensing UAVs.
	The	UAVs learn the optimal task allocation using a RL-based algorithm.
	
	We have summarized the related works and compared them with our work in Table. \ref{summary of related works}.
	\subsection{Paper Contributions}
	We aim to address the challenges of accuracy and fairness of the LSA spectrum sharing method by utilizing cooperative spectrum sensing (ELSA) and introducing a new auction model, respectively. We implement a system that utilizes reinforcement learning (RL) techniques to allocate frequency spectrum in the context of cellular networks. Our focus is on developing a new model that integrates a novel fairness-aware and dynamic auction method within the LSA framework.	
	The main contributions of this paper can be summarized as follows:\\
	\indent 1) We present a new model of an ELSA architecture, incorporating advanced technologies in telecommunications networks such as UAVs and spectrum sensing.
	Since the MNOs can simply associate the LSA infrastructures such as controllers, to their existing network and take advantage 
	from the LSA network connectivity, our objective is to address the challenges associated with an LSA-based system model as a preferred spectrum sharing method for next generation of cellular networks.\\
	\indent 2) Since we employ an auction-based spectrum allocation approach to improve the LSA method, to foster fairness in participation of MNOs, we adopt a modified VCG auction, named fair VCG (FVCG), where our preferred mechanism refines and incorporates the VCG method into the auction implementation and prioritizes fairness as a fundamental component.
	\\
	\indent 3) The proposed auction method is novel in its nature as it operates dynamically and in a repeating manner. Furthermore, the method is based on the historical operation of MNOs, as the winning of MNOs in each auction is impacted
	by their previous bids.\\
	\indent 4) Our proposed auction design is market based. This approach enables us to align the problem with real world conditions. This suggests that the success of each MNO in auctions is primarily influenced by their individual market share as well. \\
	\indent 5) In this research, we employ an innovative greedy algorithm named MSWG to determine the auction winners. Additionally, we utilize algorithms based on RL to determine \textcolor{black}{the defined weights for the bids of each MNO}. This choice has been motivated by the increasing complexity of the auctions \textcolor{black}{in real situations} and the challenges associated with allocating a growing number of demands of MNOs. As the traditional resource allocation methods are time-consuming, we incorporate the DRL-based techniques, which are instrumental in addressing this issue. Finally, we compare our innovative greedy algorithm with methods based on RL algorithms.
	\subsection{Organization of the Paper}
	The rest of this paper is organized as follows: Section \ref{III} describes the system model, assumptions and problem formulation. Section \ref{IV} presents the solution of the problem. Section \ref{V} presents the simulation results and the conclusions are stated in Section \ref{VI}.\\
	{\footnotesize{\indent $\mathbf{Symbol}$ $\mathbf{Notations:}$}} {\footnotesize{Vector and matrices are indicated by
	bold lower-case and upper-case characters, respectively. $\mathcal{A}$ denotes set $\{1,...,A\}$, and $\mathbb{R}^n$ is the set of $n$ dimension real numbers. Moreover, the notation $(\cdot)^T$ denotes the transpose operator, $|\cdot|$ indicates absolute value, $\parallel\cdot\parallel$ refers to the norm of vectors and $\otimes$ indicates the outer product of two vectors.}}
		%%%%%%%%%%%%%%%%%%%%   Table    %%%%%%%%%%%%%%%%%%%%%%%%%
	\begin{table}[h]
		\centering
		\caption{Table of Notations} \label{notations}
		\scalebox{0.7}{
			\begin{tabular}{|c|l|}
				\rowcolor{blue!20}
				\hline
				\textbf{Parameters} & \textbf{Definition} \\ \hline
				$\mathcal{M}/M/m$    &    Set/number/index of MNOs    \\\hline
				$\mathcal{M'}/M'/m'$   & Set/number/index of winning MNOs in each auction  \\\hline
				$\mathcal{K}^m/K^M/k^m$    &    Set/number/index of UAVs of $m$-th MNO    \\\hline
				$\mathcal K^{\text{Tot}}/K/k$    &    Set/number/index of all UAVs in the network    \\\hline
				$\mathcal C/C/c$    &   Set/number/index of all resource blocks of the ELSA system  \\\hline
				$\mathcal{Y}_{j,c}/Y_{j,C}/y_{j,c}$    &    Set/number/index of all binary indicators for status of the $c$-th \\  & resource block obtained form cooperative sensing for $j$-th auction    \\\hline
				$\mathcal{B}_j/b_j$   & Set/index of the bidder MNOs for $j$-th auction\\\hline
				$\mathcal{\hat B}_j/\hat b_j$   & Set/index of the bidder MNOs for $j$-th auction\\\hline
				$\mathcal{P}_j/p_j$   & Set/index of the requested resource blocks as a package \\  & of bidder MNOs for $j$-th auction\\\hline
				$\mathcal{V}_j/v_j$   & Set/index of package's value for  bidder MNOs in $j$-th auction\\\hline
				$x^{m,k}_{j,c}$   & Sensed incumbent signal in $k$-th UAV of $m$-th MNO\\  & in $j$-th auction in $c$-th resource block \\ \hline
				$s^{t}_{j,c}$   & Transmitted signal from incumbent at time slot $t$ \\  & and $j$-th auction in $c$-th resource block \\ \hline
				%$\gamma^{m,k}_c$   & SINR at the UAVs \\\hline
				$z_{j,c}^{m}$   & Fused sensing results of $m$-th MNO in $j$-th auction\\  & for $c$-th resource block \\\hline
				$C^{\text{cap}}_{j}$   &  All vacant resource blocks as the capacity of knapsack problem \\\hline
				$\varrho^{\text{Rev}}_{m',j}$   & Average selling price per unit of the resource blocks for $m'$-th \\  & MNO in $j$-th auction \\\hline
				$\mathbb{U}^{\text{Rev}}_{m',j}$  & Revenue function of $m'$-th winner in $j$-th auction \\\hline
				$\mathbb{U}_{m',j}$  & Utility function of the winning MNOs until $j$-th auction \\ \hline
				$\mathbb{U}_{m,j}$  & Utility function of $m$-th MNO until $j$-th auction \\ \hline
				$\delta^{m}_{j-1}$    & The number of requests of $m$-th MNO from the first auction\\  & to the last one $(j-1)$ \\\hline
				$\varphi^{m}_{j-1}$    &The number of requests of $m$-th MNO from the first auction \\  & to the last one $(j-1)$\\\hline
				$B_{\circ}$     &   Total available bandwidth for the LSA licensees \\\hline
				$s$   &   Bandwidth of each resource block  \\\hline
				$t_s$   & Sensing time of UAVs \\\hline
				$t_d$   & Transmission interval for winning MNOs in each auction \\\hline
				%$v$   & Speed of the UAVs \\\hline
				$H$   & Flight altitude of UAVs \\ \hline
				$D$  & UAVs trajectory period time \\\hline
				$\mathbf{r}$  & Radius of UAVs' circular trajectory  \\ \hline
				$\theta$  & Sensing angle \\ \hline
				$\alpha$  & Auction angle \\\hline
				$\lambda$   & Energy detection threshold \\\hline
				$n$   &   Sensing decision threshold\\\hline
				%$x$  &
				$\xi^{m,k}_{j,c} \in \left\{0,1\right\}$	&  Binary sensing result indicator, that if $k$-th UAV detected \\  & $c$-th resource block as occupied it is $1$, and otherwise $0$\\
				\hline
			\end{tabular}
		}
	\end{table}
	%%%%%%%%%%%%%    Table %%%%%%%%%%%%%	
	\section{SYSTEM MODEL AND PROBLEM FORMULATION}\label{III}
	\subsection{Overview}
			%%%%%%%%%%%%%%%%%%%%%  System model FIGURE  %%%%%%%%%%%%%%%%%%%%%%%%%%%
	\begin{figure*}[t]
		\centering
		\subfigure[\label{system model}]{\includegraphics[width=0.7\textwidth]{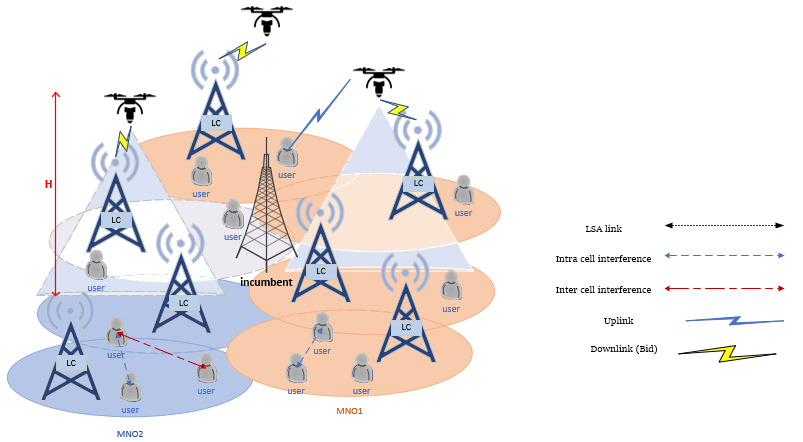}} \,\,
		\subfigure[\label{fig:sensing}]{\includegraphics[width=0.25\textwidth]{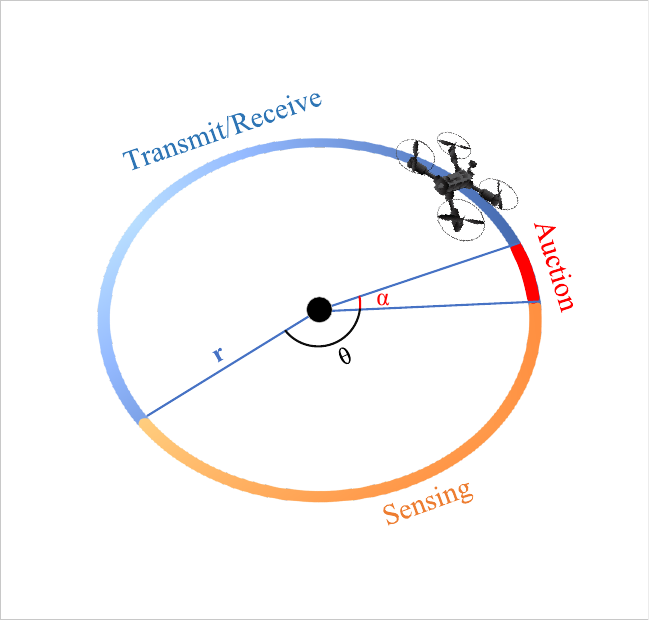}} \,\,
		\caption{\textcolor{black}{Topology of the UAV network with an incumbent base station and two MNO base stations \ref{system model}, The circular trajectory of UAV \ref{fig:sensing}, System model.}}
		\label{fig:total system model}
	\end{figure*}
			%%%%%%%%%%%% frame structure %%%%%%%%%%%%%%
	    \begin{figure}[b]
		\centering
		\includegraphics[width=1\linewidth]{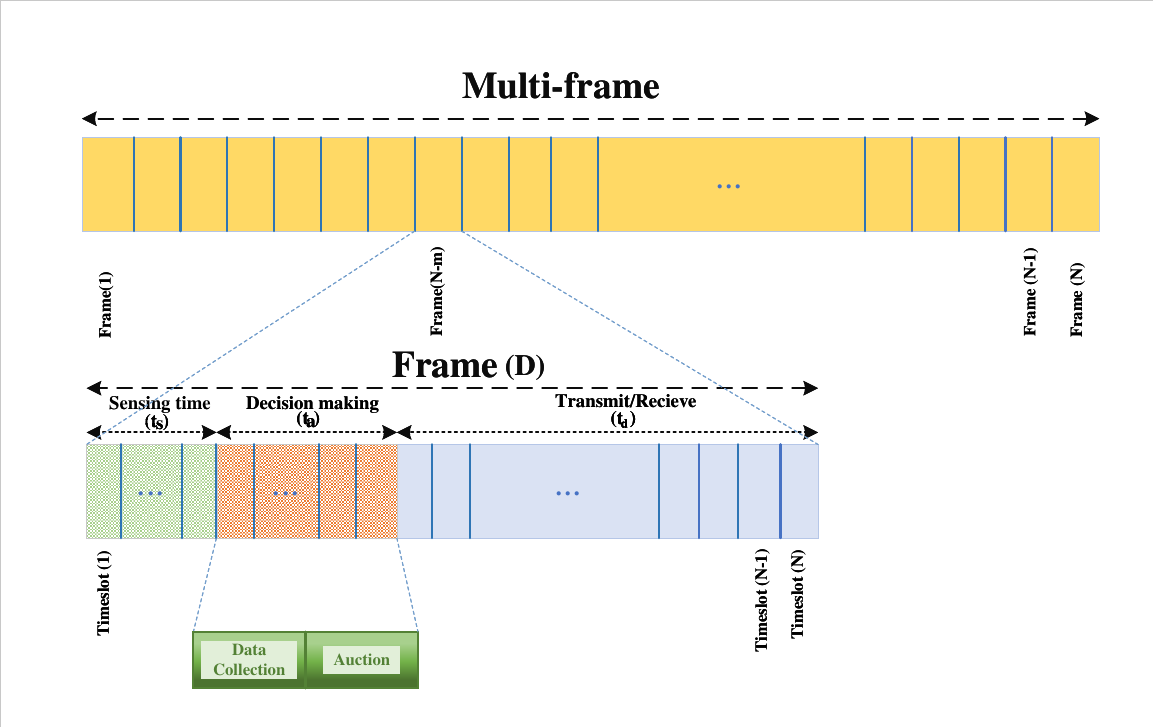}
		\caption{System operation frame structure.}
		\label{frame structure}
	\end{figure}
	As illustrated in Fig. \ref{system model}, we describe the proposed system model that has four main parts: 1) spectrum sensing with the help of UAVs by considering cooperative sensing approach, 2) spectrum auction framework design for the spectrum allocation, 3) our proposed FVCG auction structure, and 4) problem formulation. Similar to \cite{hosseini2019uav}, we primarily focus on design and implementation of spectrum sensing and a fair resource allocation, rather than concentrating on the UAV‌ wireless communication and trajectory aspect.
	 
	We assume that each of the MNOs has a number of UAVs that are spread randomly in the desired area and start their movement from a random point in a hypothetical circle with a radius of $\mathbf{r}$ with constant altitude ($H$) and velocity of $v$. It is notable that the circular UAV movement optimizes energy consumption and covers a wider area compared to other flying patterns. We assume that each UAV senses the spectrum of the incumbent in the interval $\theta$ of its periodical circular movement, sends the obtained information to the home MNO‌ and then helps the home MNO transmissions of its network in the interval $(2\pi - \theta)$, as shown in Fig. \ref{fig:sensing}. 
	The sensing results are collected by the LSA band manager (LBM) for the auction and spectrum allocation for the MNOs by auction.
    We define the set $\mathcal{M}=\left\{1,2,... ,m,...,M\right\}$ of MNOs which are within the range of the incumbent. Moreover, we consider the sets of $\mathcal{K}^{m}=\left\{1,...,K^{m} \right\}$ and $\mathcal{K}^{\text{Tot}}=\left\{\mathcal{K}^{1},\mathcal{K}^{2},...,\mathcal{K}^{M} \right\}$ to determine the number of UAVs which belong to the $m$-th MNO and the set of all UAVs, respectively. Moreover, the frame structure of this system model is shown in Fig. \ref{frame structure}.
    
\begin{figure}[b!]
		\centering
		\includegraphics[width=1\linewidth]{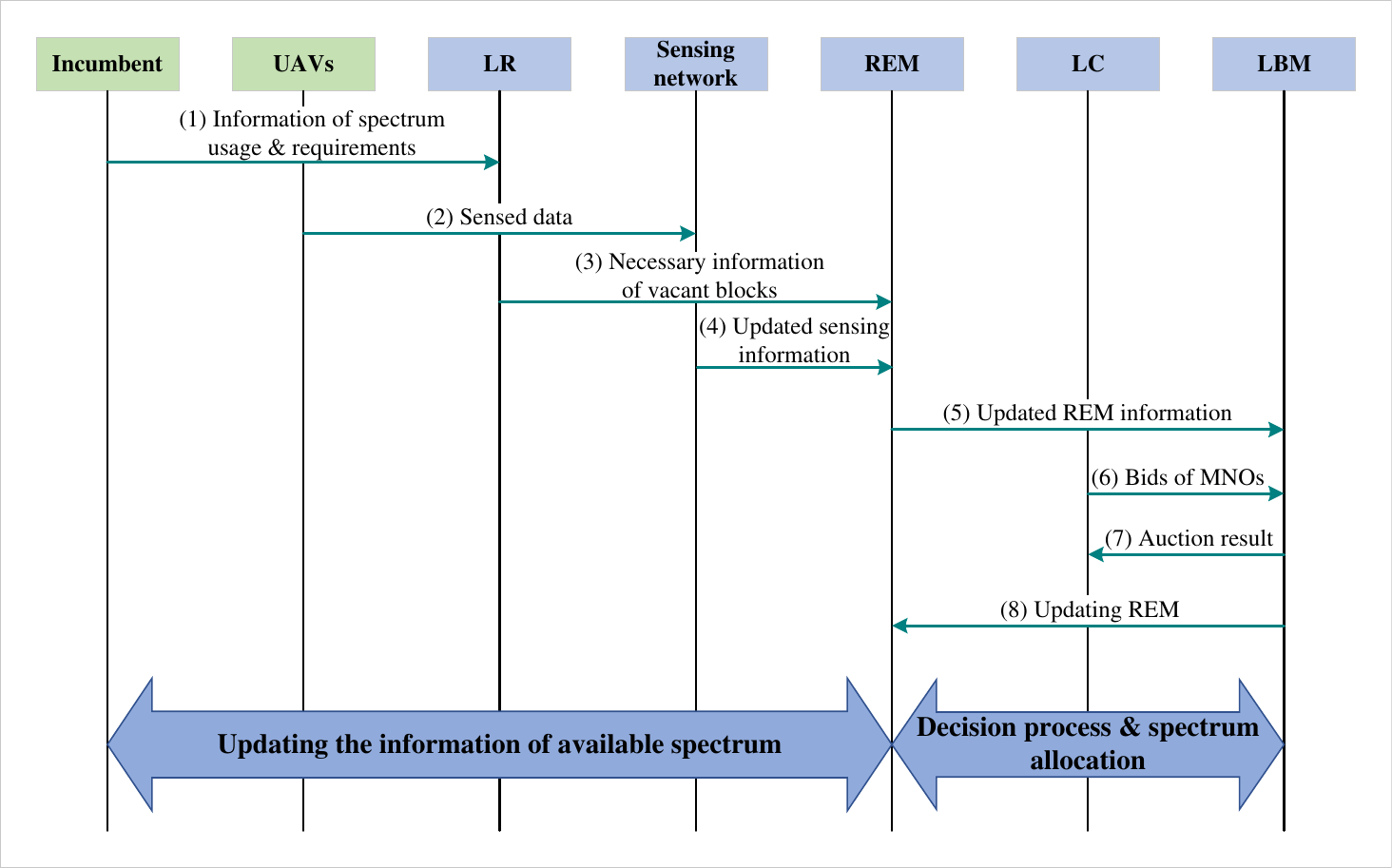}
		\caption{Call flow of the proposed architecture.}
		\label{fig:call-flow}
	\end{figure}

	$\bullet$ \textbf{LSA Mechanism:}
	Our proposed ELSA spectrum sharing method is based on the European ADEL project, which utilizes a licensed spectrum of an incumbent as a primary licensee. To improve the spectrum allocation performance, the ADEL project employs a sensing-based LSA approach \cite{frascolla2016dynamic}. In the LSA method, the LSA repository (LR) gathers information on the spectrum usage of incumbent. The LR also receives and stores acknowledgment information from the LSA controller, which manages the access of the incumbent to the available spectrum. All the sensing data collected by UAVs is sent to the radio environment map (REM) to update a map of spectrum usage in real-time. The functionality of this system is shown in Fig. \ref{fig:call-flow} as a call flow.\\
					%%%%%%%%% call flow %%%%%%%%%%%%%%%
	%%%%%%%%%%%%%%%%%%%%%%%%%%%%%%%%%%%%%%%
	When the demand of users in an area increases, an MNO sends a bid for the LBM. Due to the this system model, an auction is held based on the REM information. Spectrum sensing is performed frequently to keep the REM up to date. Ultimately, the spectrum requests and sensing data assist the LBM, who acts as an auctioneer and the manager of the system to allocate spectrum to the worthy MNOs.
		%%%%%%%%%%%%%%%%%%%   enhanced-LSA architucutre   %%%%%%%%%%%%%%%%%%%%%%%
	\begin{figure*}[t]
		\centering
		\includegraphics[width=0.7\linewidth]{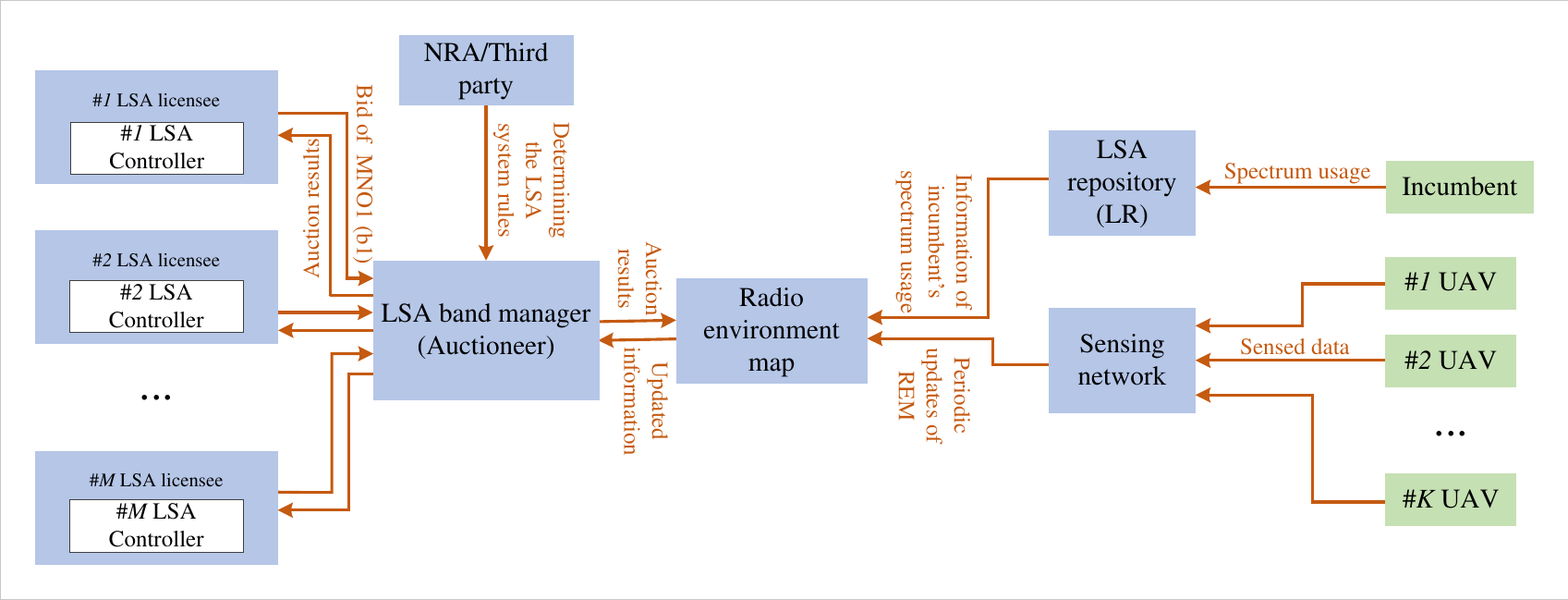}
		\caption{ELSA architecture.}
		\label{fig:enhanced-lsa-architecture}
	\end{figure*}
	As shown in Fig. \ref{fig:enhanced-lsa-architecture}, the NRA or a third party can supervise the auctions, agreement rules, and sort priorities.
\subsection{Spectrum Sensing}\label {C.} 
 In this particular scenario, we consider a circular flight of UAVs that serve as a sensing component, providing status updates of incumbent during an interval of time slots ($t_s$). We specifically choose UAVs for this purpose as they achieve superior sensing performance compared to ground spectrum sensing. Additionally, UAVs are not impacted by ground-based challenges such as multi-path fading and shadowing \cite{liu2018spectrum}. 
 Moreover, we assume that the status of incumbent does not change during a frame time interval (D). Our working assumption is that the spectrum sensing model is a cooperative central spectrum sensing approach \cite{wu2021optimisation}. In other words, each UAV independently performs spectrum sensing, and the local information of each UAV are then transmitted to the REM as a common receiver. The REM is able to collect and combine this information in order to accurately infer the presence or absence of the incumbent in each resource block. 
 
\begin{figure}[hb!]
	\centering 
	\includegraphics[width=0.6\linewidth]{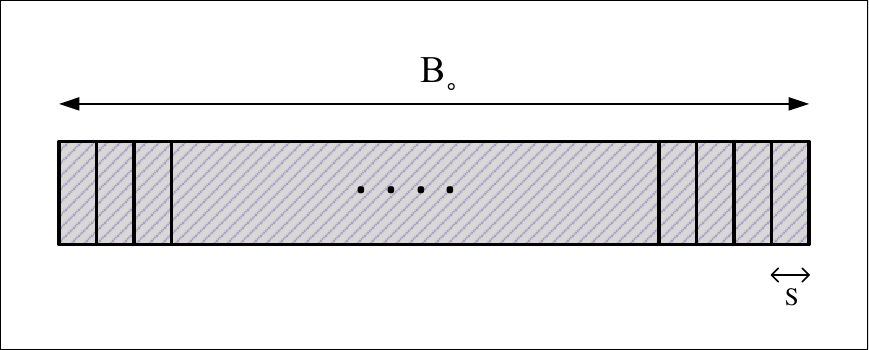}
	\caption{Spectrum partitioning based on available bandwidth for each auction.}
	\label{fig: sprectrum partitionins}
\end{figure}

 Furthermore, we represent the real status of incumbent as
\begin{align}
\begin{cases}
	H^{0}_{j,c}: & \text{incumbent is absent in the $c$-th resource} \\ & \text{block for the $j$-th auction}; \\
	H^{1}_{j,c}: & \text{incumbent is operating in the $c$-th resource} \\ & \text{block for the $j$-th auction}.
\end{cases}
\end{align}
We define a set of $\mathcal{C} = \left\{1,...,c,...C\right\}$ for all resource blocks that can either be vacant or busy by incumbent in each frame time. To divide the LSA frequency spectrum into manageable blocks, we can use a bandwidth value of $s$ for each block. Fig. \ref{fig: sprectrum partitionins} illustrates how the spectrum is divided into $C$ resource blocks so the total available bandwidth, $B_{\circ}$, determines the number of blocks which can be considered in the next auction, which is represented by the formula $C = \frac{B_{\circ}}{s}$. 
Moreover, we assign specific time slots for data collection and decision-making ($t_a$) for conducting auction $j$ and allocating the spectrum to the winning bidders. 

Furthermore, we can formulate the sensed signal of the incumbent in each sensing node as
\begin{equation}
	x^{m, k}_{j,c} = 
	\begin{cases}
	N_\circ,  & H^{0}_{j,c}\\
	h^{m, k}_{j,c}s_{j,c} + N_\circ,  & H^{1}_{j,c}
	\end{cases}
\end{equation}
where $x^{m, k}_{j,c}$ is the received signal at the $k$-th UAV of the $m$-th MNO in the $j$-th auction for the $c$-th resource block. Moreover, $h^{m, k}_{j,c}$ denotes the complex channel gain of the sensing channel between the incumbent and the $k$-th UAV, and $s_{j,c}$ is the original signal of incumbent. Assuming that the sensing time is shorter than the coherence time of the channel, we can consider the sensing channel gain $h^{m, k, t}_{j,c}$ as time-invariant during the sensing process. Hence, we can simplify the sensing channel as $h^{m,k}_{j,c}$ \cite{zhang2009optimization}. Moreover, $N_\circ$ is the additive white Gaussian channel noise (AWGN). 
%%%%%%%%%%%%%%%%%%%%%% sprectrum partitionins %%%%%%%%%%%%%%%%%%%%%%%%%
Since each UAV periodically senses the incumbent within a radius of $\mathbf{r}$ and a sensing angle of $\theta$ in its circular trajectory, the spectrum sensing time interval can be determined as
	$t_{s} = \frac{\mathbf{r}\theta}{v},$
and by considering the angle of $\alpha$ for decision making period, the transmission time interval is given by
$t_{d} = \frac{\mathbf{r}(2\pi - (\theta + \alpha))}{v}.$
We assume that the UAV utilizes energy detection to sense the spectrum of incumbent. This implies that we consider a resource block as busy if the detected energy exceeds a certain value, denoted as '$\lambda$'. The value of $\lambda$ is determined based on the permissible interference in the allowance zone, which we take from the previous studies conducted in \cite{zhang2019multi}. Therefore, the energy of the sensed signal by each UAV during $t_s$ can be defined as follows:
\begin{align}
	E(x^{m,k}_{j,c}) = |x^{m,k}_{j,c}|^{2}, \forall m\in \mathcal{M}, \forall k\in \mathcal{K}.
\end{align}
Furthermore, based on the cooperative spectrum sensing concept, each UAV makes a binary decision based on its local observation defined as a binary indicator where
\begin{equation}
	\begin{cases}
		\xi^{m,k}_{j,c}=1,  &  \text{if} \indent E(x^{m,k}_{j,c}) < \lambda,
		\\
		\xi^{m,k}_{j,c}=0,  &  \text{if} \indent E(x^{m,k}_{j,c}) \geq \lambda.
	\end{cases}
\end{equation}
In other words, $\xi^{m,k}_{j,c}=0$ if the $k$-th UAV senses the signal of incumbent and otherwise, $\xi^{m,k}_{j,c}=1$. Moreover, the REM collects all the $1$-bit decisions and fuses them together through an error-free channel according to logic rule.
The summation of all UAV binary decisions are collected in REM as the fusion center. Therefore, the binary indicator for status of the $c$-th resource block in $j$-th auction obtained from all MNOs is represented as
\begin{align}
	y_{j,c} =  \sum_{m\in\mathcal M} \sum_{\mathcal{K}^m \in \mathcal{K}^{\text{Tot}}} 	\xi^{m,k}_{j,c}
	\begin{cases}
		\geq n, & \mathcal{H}^{1}_{j,c} = 1,\\
		<n, & \mathcal{H}^{0}_{j,c} = 0,
	\end{cases}
\end{align}
where $n$ is the sensing decision threshold which is an integer, retrieved from the "$n$-out-of-$K$" voting rule \cite{banavathu2015optimal}, and $n_{\max} = |\mathcal{K^{\text{Tot}}}|$ is the total number of UAVs in the system. Moreover, we define $\mathcal{Y}_{j} = \{y_{j,c}, ... ,y_{j,C}\}$ as a tuple of binary indicators denoting all vacant resource blocks in the $j$-th auction for $c \in \mathcal{C}$. In addition, we assume that the cooperative sensing is perfect, so the vacant resource blocks are correctly deduced. Moreover, $\mathcal{H}^{1}_{j,c}$ and $\mathcal{H}^{0}_{j,c}$ denote the inference of LBM from the REM information, stating that the incumbent is operating in the $c$-th resource block or not, respectively. 

\subsection{Spectrum Auction Framework Design}\label {}
In our dynamic spectrum allocation problem, we allocate the required bandwidth to each MNO using an auction to maximize spectrum utilization. Since the auction is a transparent process, all participants have equal opportunity to bid for resource blocks. This study aims to examine VCG as a standard auction method and apply it to our problem. Therefore, we assume that the bidders are MNOs which bid for available resource blocks based on cooperative sensing results. In order to conform to the LSA framework and LTE standards, the spectrum is partitioned into discrete $s = 5$MHz resource blocks \cite{feng2014flexauc} and it is important to note that the allocated spectrum is valid for a specific duration ($t_d$). The VCG method is chosen as it is suitable for conducting auctions with one seller and multiple buyers, where each winner can obtain more than one item. This method fits very well to our system model.
 Additionally, VCG is recognized as a fair and truthful auction, which is a significant factor in this study.
 
 In our system model, multiple MNOs bid for a package of resource blocks and the LBM is the auction holder. Since the VCG auctions can be modeled as $0/1$ knapsack problem \cite{bikhchandani2001linear}, the LBM collects bids and solves the auction as a knapsack problem to choose the best combination of bids. However, our system model includes bid weights that can be updated based on the history of previous auctions to ensure the long-term fairness between all MNOs based on their market share. The following itemized paragraphs will describe the general three-stage process for conducting this VCG auction, then the role of weights in our novel auction is described in the next section.\\
\indent \textit{\textbf{1) Winner Determination:}}
 At first, we denote the set of winning MNO by $\mathcal{M'} = \{1,...,m',...,M'\}$ where $\mathcal{M'} \subseteq \mathcal{M}$. 
In a rational market, each MNO aims to maximize its profits by submitting considerable bids in each auction, taking into account its demand, urgency and budget. For the winner determination, the following approach is considered: In the first step, bids are collected from all MNOs. In order to solve the knapsack problem, the LBM considers all possible combinations of bids and places them into a set. 
The largest element of the combination set is then selected, in such a way that if an MNO wins, it receives the entire package which was offered, and otherwise, no spectrum block is assigned to it. It should be noted that as the number of bidders and spectrum blocks increase, the process of solving this problem becomes more complex. Moreover, we consider $\mathcal{B}_{j} = \left\{b^{1}_{j}, \cdots,b^{m}_{j},\cdots,b^M_j \right\}$, $\mathcal{P}_{j} = \left\{p^{1}_{j},\cdots, p^{m}_{j},\cdots,p^{M}_{j}\right\}$, $\mathcal{V}_{j} = \left\{v^{1}_{j},\cdots, v^{m}_{j},\cdots,v^{M}_{j}\right\}$ as the set of bids, the set of the requested spectrum blocks as a package where the elements of this set should be less than or equal to the number of available resource blocks, and the value of requested package for each MNO, respectively. The capacity of knapsack problem in the $j$-th auction is equal to vacant resource blocks obtained from the LBM sensing decision, and this is expressed mathematically as
$C^{\text{cap}}_{j} = \sum_{c \in C}y_{j,c}.$

Therefore, an auction is held when the number of requests for spectrum blocks in all packages exceeds the network capacity, i.e., when
	\begin{align}
	 \label{C1}& \text{C1} : \sum_{m\in \mathcal{M}}{p^{m}_{j}} > C^\text{cap}_{j}, \forall m \in \mathcal{M}, j \geq 1.
\end{align} 
Otherwise, if this condition is not met, the auction does not proceed, and the resource blocks are allocated to the existing bidders in the $j$-th auction. 
Our other assumption is that the ordering of the blocks is not a significant factor, and only the quantity of resource blocks contained in each package is taken into consideration. So in the execution of the algorithm, we only count the number of resource blocks. We make the additional assumption that if the $m$-th MNO has not submitted any request, then the values of $b^{m}_{j}$, $p^{m}_{j}$ and $v^{m}_{j}$ variables will be equal to zero.  According to the assumptions raised in the VCG auction, the MNOs shall lack the knowledge about the bids  and values of one another.
\\
\indent \textit{\textbf{2) Payment Mechanism:}}
 \normalsize The payment price of each winner is equivalent to its social opportunity cost, which represents the externality imposed by the \textcolor{black}{winning} MNO on other bidders in the auction. Therefore, we consider two concepts:\\
 $\bullet$ \textbf{Original auction} which consists of all $M$ MNOs as auction participants, whose social value or total bid is denoted by  $B_{\mathcal{M}\diagdown\left\{-m'\right\}}$ in formula (\ref{winner cost function}). It is calculated by removing the $m'$-th MNO from the auction and determining the best combination of bids.\\
 $\bullet$ \textbf{Modified auction} which involves none of the winning MNOs and its social value,  $B_{\mathcal{M}}^{-b_{m'}}$, is obtained by excluding the bid of the $m'$-th MNO from the best combination of bids in the original auction and calculating the sum of remaining bids. Hence, the cost function which is paid by the $m'$-th winning MNO in the $j$-th auction, is expressed as follows:
\begin{equation} \label{winner cost function}
	\mathbb{U}^{\text{Cost}}_{m',j} = 
	B_{\mathcal{M}\diagdown\left\{-m'\right\}} - B_{\mathcal{M}}^{-b_{m'}}, \forall m'\in \mathcal{M'},
\end{equation}
\indent \textit{\textbf{3) Revenue Function Model:}}
\normalsize
We let $\varrho^{\text{Rev}}_{m',j}$ represent the average selling price per unit of the resource blocks for the $m'$-th MNO as a winner in the $j$-th auction, whose unit is \$/Mbps. Then, the revenue function of the winning MNOs is defined as
\begin{equation} \label{winner revenue function}
	\mathbb{U}^{\text{Rev}}_{m',j} =  \sum_{m'\in M}{p^{m'}_{j}}{\varrho^{\text{Rev}}_{m',j}}, \forall m'\in \mathcal{M'},
\end{equation}
and the utility function of the winning MNOs is defined as
\begin{align}
	\mathbb{U}_{m',j} = \mathbb{U}^{\text{Rev}}_{m',j}-	\mathbb{U}^{\text{Cost}}_{m',j}, \forall m'\in \mathcal{M'},
\end{align}  
where the $\mathbb{U}^{\text{Rev}}_{m',j}$ is the revenue of the $m^\prime$-th winning MNO from its users service, and $\mathbb{U}^{\text{Cost}}_{m',j}$ is the winner price of the $m'$-th winning MNO in the $j$-th auction, where we assume that this value is fixed for all spectrum blocks.
Moreover, the utility function of each MNO until the $j$-th auction is defined as
\begin{align} \label{total utility function}
	\mathbb{U}_{m,j} = \sum_{i=1}^{j}{\mathbb{U}_{m',j}}, \forall m'\in \mathcal{M'}, m=m',
\end{align}
where the set of $\mathbb{U}_j = \left\{\mathbb{U}_{1,j}, \cdots, \mathbb{U}_{m,j},\cdots, \mathbb{U}_{M,j}\right\}$ is the utility of all MNOs in the $j$-th auction.
\subsection{FVCG Auction} 
Due to the inability of the auction mechanism to guarantee the fairness of winning among all bidders and the requirement of our system model to perform the auction
continuously, we propose a novel weighted mechanism, named FVCG auction, to achieve the fairness from LBM point of view. In particular, the proposed FVCG auction takes into account not only the value of bids but also the market share of each MNO and its previous auctions history. By considering these additional parameters, we aim to increase the fairness index of the ELSA system over time.

The fairness is quantified based on equity theory \cite{carrell1978equity}. In equity theory, one evaluates fairness by comparing effort and outcome.  In this study, fairness refers to the goal of creating equal opportunities for all MNOs to win in different auctions, regardless of the number of requests they make. To ensure fairness in the market, we check the fairness of the market every few auctions and adjust the weights of the bids accordingly to prevent any monopolization in the market. Thus, we expect that after some auctions, fairness index of the network converges to an appropriate value, so that all MNOs who request resource blocks will have an equal chance of winning in the long-term. To achieve this, the auction takes into account the market share of each MNO by specific primary coefficients to bids.
Let $w^{m}_j$, $\delta^{m}_{j-1}$ and $\varphi^{m}_{j-1}$ represent the updateable weight of bid, the number of previous requests, and the number of previous winnings from the first auction to the $(j-1)$-th auction of the $m$-th MNO, respectively. So, if we update the weights every $x$ auctions, it means that if $\frac{j-1}{x} \notin \mathbb{N}$ we have $w^{m}_j = w_{j-1}^m$ and otherwise, we \textcolor{black}{
will update the} weights. The following itemized paragraphs describe the different weights definitions inspired by \cite{pla2015multi}:\\
\indent \textit{\textbf{1) Win per request-based:}} 
In this case, one of the weights which is used to affect the allocation of MNOs, is based on the variables $\delta^{m}_{j-1}$ and $\varphi^{m}_{j-1}$. The rationale behind selecting this weight is to assign a higher weight to MNOs that despite their requests, they did not win in previous auctions. By doing so, we aim to increase their chances of winning in the upcoming $x$ auctions. Nevertheless, our goal is to enhance fairness within the network by aligning the probability of MNOs winning with their respective market share and activities. Thus, we define the relevant weight as follows:
\begin{align} \label{Weight Based on Wins and Requests}
	 \hat{w}^m_{j} = 1 - \left(\frac{1 + (\delta^{m}_{j-1})}{1 + (\varphi^{m}_{j-1})}\right).
\end{align} 
\indent \textit{\textbf{2) Utility-based:}}
%\indent $\bullet$ \textbf{Utility-based:}
In this case, the weights assigned to MNOs are determined based on their utility generated by each MNO from the first auction until present. As represented in \eqref{winner revenue function}, \textcolor{black}{
 the requested package of each MNO depends on the demand of its users}. \textcolor{black}{Hence, it} serves as an incentive for MNOs to offer the highest possible bid.
Since achieving "truthfulness" is crucial in any auction, and the bids of the MNOs are influenced by the value of resource blocks, the following weight is defined in such a way that it encourages MNOs to bid truthfully, i.e., the weight guarantees that players report their true values. Mathematically, this weight is defined as follows:
\begin{align}  \label{Weight Based on utility}
	\Hat{\Hat{w}}^m_{j} = 1 - \left(\frac{\sum \limits_{i = 1}^{j-1}(\mathbb{U}_{m,i})}{\sum \limits_{m\in \mathcal{M}} \sum \limits_{i = 1}^{j-1}(\mathbb{U}_{m,i})}\right).
\end{align}
\indent \textit{\textbf{3) Combined Weight:}}
%\indent $\bullet$ \textbf{Combinatorial:}
The following weight is derived by multiplying the two preceding weights, to contain all the used parameters. The formula can be expressed as follows:
\begin{align} \label{Combined weight}
	w^m_{j} = \hat{w}^m_{j}.\Hat{\Hat{w}}^m_{j}.
\end{align}
Additionally, we can define $\mathbf{b}_{j}$ and $\mathbf{w}_{j}$ as two vectors whose components are the bids and weights of each MNO in the $j$-th auction, respectively as follows:
\begin{align}
	&\mathbf{b}_{j} =  (b^{1}_{j}, \cdots , b^{m}_{j}, \cdots, b^{M}_{j}) , \forall m \in \mathcal{M},  \\
	&\mathbf{w}_{j} = (w^m_{j}, \cdots, w^m_{j}, \cdots, w^M_{j}), \forall m \in \mathcal{M},
\end{align} 
where the updated bids' matrix for the $j$-th auction is obtained as the result of the outer product
	$\mathbf{\hat B}_{j_{m \times m}} = \mathbf{w}_{j} \otimes (\mathbf{b}_{j})^{T}.$

Therefore, we define the set of input bids of the knapsack algorithm in a set represented as
	$\mathcal{\hat B}_{j}=\left\{\hat b_{j_{11}}, \hat b_{j_{22}}, ...,\hat b_{j_{mm}}\right\}, \forall m \in M.$

It is noteworthy that at the beginning of the auctions, if $j \leq x$, the value of each element in the vector $\mathbf{w}_{j}$ is equal to $1$. In other words, the inputs of knapsack are equal to the received bids, and thus $\mathcal{\hat B} = \mathcal{B}$.

Jain's fairness index is a mathematical expression that quantifies the level of fairness in an auction. Accordingly, the fairness index is defined as follows:
\begin{align}
	f_{j} =\frac{\left(\sum\limits_{m\in \mathcal{M}} (\frac{\varphi^{m}_{j-1}}{\delta^{m}_{j-1}})\right)^{2}}{M.\sum\limits_{m\in \mathcal{M}} {(\frac{\varphi^{m}_{j-1}}{\delta^{m}_{j-1}})^{2}}}, \forall m \in \mathcal{M}, j \geq 1 .
\end{align}
This index is defined based on the win to request ratio of the MNOs, and its value ranges from $0$ to $1$. A value of $1$ indicates perfect fairness, while a value of $0$ indicates the opposite. The fairness index is used to evaluate the degree of the long-term fairness in the $j$-th auction.
\subsection{Problem Formulation}
Based on these definitions, our aim is to solve the following optimization problem:
\begin{subequations}
    \begin{align}{\label{optimization}}
	    &\max_{\mathbf{w}} \sum_{j = 1}^{J}{f_j} \\
	    \mathbf{s.t.} \indent  &\text{C1} \\
		\label{C2}& \text{C2} : (1-\frac{\varphi^{m}_{j-1}}{\delta^{m}_{j-1}})>0, \forall m \in M,  j \geq 1,\\
	    \label{C3}&\text{C3} : \xi^{m,k}_{j,c} \in \left\{0,1\right\}, \forall m \in \mathcal{M}, \forall k \in \mathcal{K}, \notag \\ & \indent \indent  \forall c \in C, \forall j \geq 1,\\
	    \label{C4}&\text{C4} : y_{j,c} \in \left\{0,1\right\},  \forall c \in C, \forall j \geq 1.
    \end{align} 
\end{subequations} 
 where $J$ is the auction number
 we are in, at the time of calculating the fairness index and the details of constraint C1 is already provided in \eqref{C1}. The constraint C2 is used to ensure that the weights are positive and non-zero. In addition, constraint C3 and C4 are used to ensure that sensing decision variables are binary. 
\section{Optimization Problem Solution}\label{IV}
\subsection{Preliminary Discussion}
The problem formulation (\ref{optimization}) presented in this study is a challenging non-convex, mixed-integer nonlinear programming (MINLP) problem that poses difficulties in achieving a global solution for a large network. Traditionally, VCG auctions were solved by the knapsack algorithm, a greedy algorithm that would not result in a high-fairness network. Therefore, in order to address this problem, we apply two distinct approaches. In the first approach, we solve our proposed novel FVCG auction with the help of an improved greedy algorithm named market share-based weighted greedy algorithm (MSWGA). 
As the second approach, we leverage two online DRL-based algorithms for handling resource allocation to reach market share-based fairness of the network. This requires reformulating the problem as a Markov decision process (MDP) and subsequently employing deep deterministic policy gradient (DDPG) and soft actor-critic (SAC) algorithms for its solution.
\subsection{Market Share-based Weighted Greedy Algorithm (MSWGA)}
%%%%%%%%%%%%%%%% flow chart figure %%%%%%%%%%%%%%%%%%%%%%
\begin{figure}[b]
	\centering 
	\includegraphics[width=0.8\linewidth]{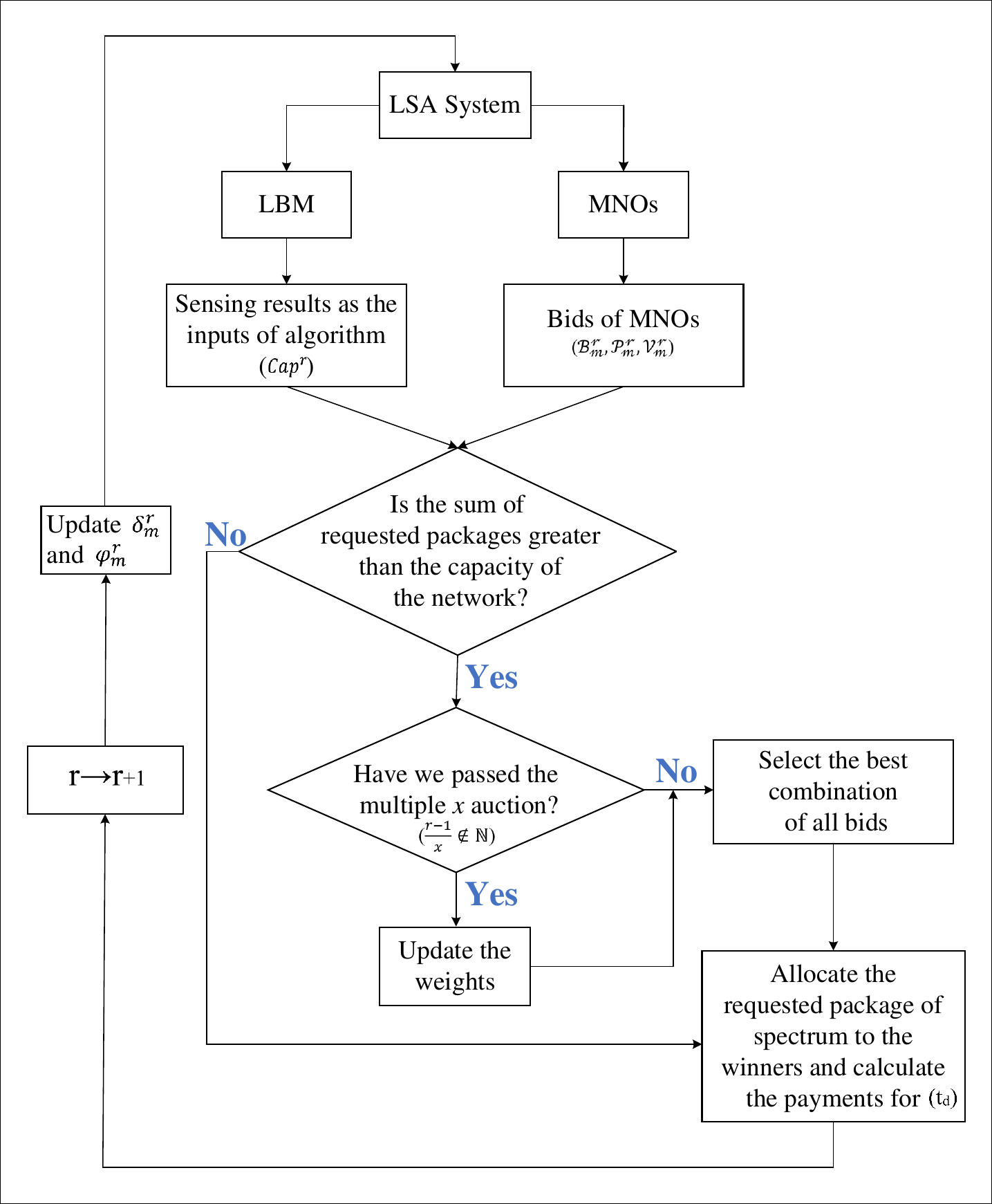}
	\caption{\color{black}{The flowchart of MSWGA algorithm.}}
	\label{MSWGA}
\end{figure}
%%%%%%%%%%%%%%%%% flwo chart %%%%%%%%%%%%%%%%%%%%%%
The winner determination in the VCG auction is formulated as a knapsack problem, which is used for the problems with limited capacity. It is solved using a greedy algorithm as a dynamic programming approach. Based on traditional knapsack solution, at $j$-th auction, the greedy algorithm picks the option that is locally optimal, meaning that it looks like the most suitable option right now and the winner determination depends on the other bids of $j$-th auction. In the MSWGA, the parameter for establishing fairness in the auction process involves the calculation of a ratio between the number of times an MNO wins and the number of its requests from the first auction to $(j-1)$-th one. To ensure fairness, these weights are periodically reassessed after $x$ stages of the auction. \textcolor{black}{
It means that in the subsequent $x$ auctions, the winning probability of certain MNOs is increased by multiplying specific weights with the bids of those MNOs, if their win to request ratio was lower}. This process is presented in Fig. \ref{MSWGA}. Additionally, at the start of this algorithm, some fixed coefficients are assigned to each MNO based on their market share ratio. For instance, if MNO $1$ has twice the market share compared to MNO $2$, a constant coefficient which is twice, is assigned to MNO $1$. This approach allows us to effectively examine the impact of auction weights on the fairness of the system. 
\subsection{DRL-based Algorithms}
A single type of agent is considered (the LBM) which selects its own actions to maximize the fairness index of the ELSA network by obtaining the \textcolor{black}{weights of bids}. The agent progressively learns a model-free policy that maps observations to optimal actions. In the DRL framework, MDP is a tuple form $\left\{\mathcal{S}_t,\mathcal{A}_t,\mathcal{R}_t,\mathcal{S}_{t+1}\right\}$ containing states, actions, rewards, and new states. The agent observes the current state of the environment, $\mathcal{S}_t$, and the corresponding reward value, $\mathcal{R}_t$, at some arbitrary time step, $t$. Based on this information, the agent decides what to do next. As a result of the action the agent intends to execute, $\mathcal{A}_t$, we can obtain the new state $\mathcal{S}_{t+1}$. Since in our problem, each time step \textcolor{black}{consists of} an auction event, the $t$ and $j$ indices are interchangeable.\\
\indent $\bullet$ \textbf{System States:} The system state is an abstraction of the environment, defined as $\mathcal{S}_{j}$. Also, $s_j \in \mathcal{S}_{j}$ represents the state of the environment at $j$-th auction, which includes the collection of bids, requested packages, and values of MNOs, along with the number of vacant resource blocks as the capacity of system, \textcolor{black}{the number of wins and requests, and the utility of MNOs. It that can be expressed as}
\begin{align}
	s_j = \left\{b^m_{j}, p^m_{j}, v^m_{j}, \mathcal{Y}_j, \delta^{m}_{j}, \varphi^{m}_{j}, \mathbb{U}_{j}^m\right\}.
\end{align}
\indent $\bullet$ \textbf{Action Space:} The agent selects an action by evaluating the current state of the network, as the network transitions to a new state from its current state. The set of action space is defined as $\mathcal{A}_{j}$, and $a_j \in \mathcal{A}_{j}$ executed in the $j$-th auction is expressed as
	$a_j = \left\{\mathbf{w}_{j}\right\},$
where $\mathbf{w}_{j}$ represents the weight assigned to the bids of each MNO. This weight is carefully determined to ensure a fair network and \textcolor{black}{is based on the history of previous auctions and the market share of MNOs.} Additionally, the corresponding weights influence the actions of states mentioned earlier in the $j$-th auction.\\
\indent $\bullet$ \textbf{Reward Function:} After an action is executed, agent receives an immediate reward, denoted as $r_j$, from the environment. In this system, the agent aims to enhance the optimization problem (\ref{optimization}), by maximizing the fairness index.  The reward represents the benefit derived from the action. Thus, the reward is denoted as
	$r_j = f_j,$ where $j \geq 1.$

\indent $\bullet$ \textbf{Discount Factor:} The cumulative long-term reward of a policy is represented by the following model
	$R_j = \sum_{i =0}^{\infty} \gamma^{i}\left(r_{j+i}\right),$
where the discount factor $\gamma \in \left[0,1\right]$ indicates the level of importance the agent places on long-term rewards. A higher value of $\gamma$ (approaching $1$) implies that the agent is more focused on maximizing long-term rewards.

Our optimization objective is to maximize the expected long-term discounted cumulative reward, expressed as
\begin{align}
	J(\pi) = \mathbb{E}_{(s_{j}, a_{j})\sim\rho^\pi}\left[\sum_{i =0}^{\infty} \gamma^{i}\left(r_{j+i}\right)\right],
\end{align}
where $\rho^\pi$ denotes the state-action distribution induced by policy $\pi$. The optimal policy, denoted as $\pi^\star$, can be learned via maximizing the expected long-term discounted cumulative reward, expressed as
		$\pi^\star = \arg \max_{\pi} J(\pi).$
\subsubsection{DDPG Approach}
The DDPG method is an off-policy and model-free RL approach that has been well-suited for handling large and continuous state and action spaces. This method, based on actor-critic architectures, utilizes deep neural networks (DNNs) as function approximators to define deterministic policies that map high-dimensional discrete or continuous states to continuous actions. In this framework an agent interacts with the environment through discrete auctions. \textcolor{black}{At each auction $j$, the LBM as agent takes an action based on the observation and receives the scalar reward $r_j$}. The DDPG method utilizes two networks: the actor network (policy $\pi$) and the critic network (value function $Q$), characterized by their respective parameters $\theta^\pi$ and $\theta^{Q}$. Additionally, two copies of the actor and critic networks are presented as target actor network and target critic network with the parameters of $\theta^{\pi^\prime}$ and $\theta^{Q^\prime}$, respectively. The $Q$-function undergoes updates using temporal-difference methods, which closely resembles the approach employed by DQN. The policy gradient algorithm is utilized to update the actor's value, utilizing information obtained from the critic. The state's return is calculated as the accumulation of future rewards, discounted according to a certain factor. 
Additionally, the purpose is to obtain a policy that maximizes the expected return. The Bellman equation is utilized to learn the action-value function $Q\left(s,a|\theta^Q\right)$. 
\textcolor{black}
{To compute the target values, we generate copies of the actor and critic networks, denoted as $\pi^\prime\left(s|\theta^{\pi^\prime}\right)$ and $Q^{\prime}\left(s,a|\theta^{Q^\prime}\right)$, respectively. The optimization of the objective through the utilization of the difference between the critic and target critic networks can be expressed by employing the subsequent loss function
	\begin{align}{\label{loss-function-critic}}
		L(\theta^Q) = \mathbb{E}_{s_j\sim\rho^\pi,a_j\sim\pi,r_j\sim E}\left[\left(Q(s_j,a_j|\theta^Q) - y_j\right)^2\right],
	\end{align}
	where $y_{j}$ is
	\begin{align}{\label{Q-value}}
		y_j = r_{j}+\gamma Q(s_{j+1},\pi(s_{j+1})|\theta^Q).
	\end{align}
} 
Hence, the distinction between the actor and target actor networks can be denoted by employing the subsequent loss function
\begin{align}{\label{loss-function-actor}}
	\resizebox{\linewidth}{!}
	{$L(\theta^\pi)= \mathbb{E}_{s_{j+1}\sim(s_j,\rho^\pi),a_{j+1}\sim\pi,r_j\sim E} \left[\max Q'(s_{i+1},a_{i+1},\theta^{\pi^\prime}) - Q(s_j,a_j,\theta^\pi)\right] + r_j$}.
\end{align}
Therefore, based on Bellman equation, the actor and critic networks are updated by uniformly sampling a minibatch from the replay buffer at specific episodes. Finally, the weights of target networks will be updated as
\begin{align}{\label{up-actor}}
	\theta^{Q^\prime} \leftarrow \tau\theta^Q + (1-\tau)\theta^{Q^\prime},
\end{align}
\begin{align}{\label{up-critic}}
	\theta^{\pi^\prime} \leftarrow \tau\theta^\pi + (1-\tau)\theta^{\pi^\prime},
\end{align}
where $\tau$ is the target network update period and $\tau \ll 1$. Consequently, to establish an exploration policy $\pi^\prime$, \textcolor{black}{the noise component, sampled from a noise process denoted by $\sigma$, is incorporated into the actor policy based on the environment.} Hence, the following relationship is established
\begin{align}
	\pi^\prime(s_j) = \pi(s_j|\theta_j^\pi)+\sigma_j.
\end{align}
The complete pseudo-code of the DDPG approach is represented in Algorithm \ref{DDPG}.
%****************DDPG Pseudocode*****************
\begin{algorithm}[t]
	\tiny
	\caption{DDPG Algorithm: $B$ and $D$ are the batch size and mini-batch size, respectively; $M$ is the memory counter, and $\alpha$, $\beta$ are the learning rates of actor and critic networks.}\label{DDPG}
	\textbf{Input:} Initialize parameters $\theta^Q$ of critic network $Q$ and parameters $\theta^\pi$ of actor network $\pi$; target networks $\pi^\prime$ and $Q^\prime$; set $M=0$.\\
	Receive initial states\\
	\For{\textup{episode} $j=1$ to $E$}{
		Observe $s_{j}$ and select action $a_j = \pi(s_j|\theta^\pi) + \sigma_j$\\
		Execute actions and observe reward $r_{j}$ and new states $s_{j+1}$\\
		Store  $(s_j,a_j,r_j,s_{j+1})$, in reply buffer $\mathcal{D}$ and $s_{j}\leftarrow s_{j+1}$\\
		$M \leftarrow M+1$\\
		\If {$M >= B$}{Sample minibatch of size $D$, from replay buffer $\mathcal{D}$ \\
			Set $y_j$ by (\ref{Q-value})\\
			Calculate the loss by (\ref{loss-function-critic})\\
			Update the critic by Minimizing the loss by (\ref{loss-function-critic}) \\
			Update parameters $\theta^\pi$ and $\theta^Q$ by minimizing the loss functions in (\ref{loss-function-critic}), (\ref{loss-function-actor}), using:\\
			$\theta^{\pi}_{j+1} = \theta^{\pi}_{j}-\alpha \nabla {L}(\theta^{\pi}_{j})$\\
			$\theta^{Q}_{j+1} = \theta^{Q}_{j}-\beta \nabla {L}(\theta^{Q}_{j})$\\}
		\If{j \textup{mod} $J_{\textup{up}}=1$}{
			Update parameters $\theta^{\pi^\prime}$ and $\theta^{Q^\prime}$ by (\ref{up-actor}), (\ref{up-critic})\\}
	}
\end{algorithm}
\subsubsection{SAC Approach}
There is a trade-off between exploration and exploitation in RL, in which an agent must balance exploring its environment to discover optimal policies with exploitation to maximize cumulative rewards. To address this challenge, soft actor-critic (SAC) has emerged as a powerful method, which combines actor-critic and entropy-regularized RL techniques. In addition to its capability to handle continuous action spaces, it also maintains a stable training environment and delivers high sample efficiency. Through entropy regularization, SAC maximizes both policy and value functions simultaneously, enabling exploration without sacrificing determinism.
In what follows, we explore the formulation of the SAC algorithm through a policy iteration approach. Unlike traditional methods, SAC is an off-policy actor-critic DRL method. 
Rather than solely maximizing the discounted cumulative reward, the optimal policy in SAC seeks to maximize the entropy-regularized reward. This means that the actor aims to maximize both the expected reward and the entropy of the policy. We adopt a maximum entropy RL algorithm as follows:
\begin{equation}
	J(\pi)=\sum_{j=0}^E \mathbb{E}_{\left(s_j, a_j\right) \sim \rho_\mu}\left[r_j+\varphi \mathcal{H}\left(\pi\left(\cdot \mid s_j\right)\right)\right],
\end{equation}
where $\varphi$ is the regularization coefficient which ensures that entropies and the sum of expected rewards remain finite. Moreover,  $\mathcal{H}(\pi(.|s_j))=\mathbb{E}_{a_j}[-\log\pi(a_j|s_j)]$ is the policy
entropy. The $Q$ function is learned by minimizing the soft Bellman residual, as follows:
\begin{equation}
	\begin{aligned}
		& J_Q(\theta^Q)= \\
		& \mathbb{E}\left[\left(Q\left(s_j, a_j\right)-r_j-\gamma \mathbb{E}_{s_{j+1}}\left[V_{\theta^{Q'}}\left(s_{j+1}\right)\right]\right)^2\right], \\
		& V_{\theta^{Q'}}(s)= \mathbb{E}_{\pi_{\theta ^ \pi}}\left[Q_{\theta^{Q'}}(s, a)-\varphi \log \pi_{\theta^\pi}(a \mid s)\right].
	\end{aligned}
\end{equation}
By minimizing the expected KL-divergence, the policy $\pi_\theta$ can be learned
\begin{equation}
	\label{54}
	J_\pi(\theta^\pi)=\mathbb{E}_{s \sim \mathcal{B}}\left[\mathbb{E}_{a \sim \pi_{\theta} }\left[\varphi \log \pi_{\theta}(a \mid s)-Q_{\theta^Q}(s, a)\right]\right],
\end{equation}
where $\mathcal{B}$ represents the set of previously sampled states and actions, stored in a replay buffer. To optimize $J_\pi(\theta^\pi)$, we use a likelihood ratio gradient estimator, which eliminates the need for backpropagating gradients through the target density networks and the policy. \textcolor{black}{In the case of SAC, where the target density is the Q-function represented by a neural network, we reparameterize the policy $\pi_{\theta^\pi}$ using a neural network. This network takes both the state $s$ and a noise vector $\epsilon$ as inputs, resulting in a lower variance estimator}
\begin{align}
	\label{55}
	a=f_{\theta^\pi}(s, \epsilon).
\end{align}
Accordingly, by substituting \eqref{55} in \eqref{54} we have
\begin{equation}
	\begin{aligned}
		& J_\pi(\theta^\pi)= \\
		&\mathbb{E}_{s \sim \mathcal{B}, \epsilon \sim \mathcal{N}}\left[\varphi \log \pi_{\theta^\pi}\left(f_{\theta^\pi}(s, \epsilon) \mid s\right)-Q_{\theta^Q}\left(s, f_{\theta^\pi}(s, \epsilon)\right)\right],
	\end{aligned}
\end{equation}
where $\mathcal{N}$ represents a standard Gaussian distribution, and $\pi_{\theta^\pi}$ is implicitly defined in terms of $f_{\theta^\pi}$. Lastly, the SAC updates the regularization coefficient $\varphi$ by minimizing the loss function, as follows:
\begin{equation}
	J(\varphi)=\mathbb{E}_{a \sim {\pi _{\theta ^\pi} }}\left[-\varphi \log \pi_{\theta^\pi} (a \mid s)-\varphi \mathfrak{e}\right],
\end{equation}
where $\mathfrak{e}$ represents the target entropy as a hyper-parameter. The complete  pseudo-code of the SAC approach is given in Algorithm \ref{SAC}.
%*************** SAC algorithm ******************
\begin{algorithm}[t]
	\tiny
	\caption{Soft Actor Critic (SAC) Algorithm}\label{SAC}
	\SetAlgoLined
	{{\textbf {Hyper-parameters}}: Step sizes $\lambda_\pi$, $\lambda_Q$, $\lambda_\phi$, target entropy $\varrho$, exponentially moving average coefficient $\tau$}\\
	\KwIn{Initial $Q$ value function parametes $\theta_1$ and $\theta_2$}
	\KwIn{Initial policy parameter $\mu$}
	$\mathcal{B}$ = $\emptyset$; $\theta^\prime$ = $\theta$, for $i \in \left\{1,2\right\}$\\
	\For{each iteration}{
		$a_j\sim \pi_\mu\left(.|s_j\right)$\\
		Consider the action $a_j \in \mathcal{A}_j$ as the optimization variables of problem \ref{optimization}\\
		$s_{j+1}\sim p(s_{j+1}|s_j,a_j)$\\
		Calculate the reward\\
		$\mathcal{B}$ $\leftarrow$ $\mathcal{B}$ $\bigcup$ $\left\{s_j,a_j,r_j,s_{j+1}\right\}$\\
		$\mu \leftarrow - \lambda_\pi \nabla_\mu J_\pi(\mu) $\\
		$\theta_i \leftarrow \theta_i - \lambda_Q \nabla J_Q(\theta_i)$ for $i \in \left\{{1,2}\right\}$\\
		$\phi \leftarrow \phi - \lambda_\phi \nabla J(\phi)$\\
		$\theta^\prime_i \leftarrow \tau \theta_i^\prime + (1-\tau)\theta_i$ for $i \in \left\{1,2\right\}$
	}
\end{algorithm}
%************** SAC algorithm*****************
\section{SIMULATION RESULTS}\label{V}	
\subsection{Parameter Setting}
We consider a CR system with an incumbent as primary and $5$ MNOs as secondary users
\footnote{\textcolor{black}{The implementation of our algorithm is available online at}\\
\indent \indent \href{https://ieee-dataport.org/documents/novel-dynamic-fairness-aware-auction-enhanced-licensed-shared-access-6g-networks}{ieee-dataport.org} \url{}\cite{c75s-h743-23}.}. 
Additionally, there is one UAV per MNO and the total available bandwidth for all MNOs is $100$ MHz, which has been divided into $20$ resource blocks with a bandwidth of $5$ MHz. The sensing decision threshold for detecting the presence of the primary incumbent is set at $n = 3$.
\begin{table}[h!]
	\centering
	\caption {Simulation Parameters}
	\label{chap5:Simulation parameters}
	\scalebox{0.7}{
		\begin{tabular}{|c|l|c|}
			\rowcolor{blue!50}
			\hline
			\textbf{Parameter} & \multicolumn{1}{c|}{\textbf{Description}} & \textbf{Value} \\ \hline
			\rowcolor{blue!20}
			\multicolumn{3}{|c|}{\textbf{Simulation Environment Parameters}} \\ \hline
			$M$ & Total number of MNOs & 5 \\ \hline
			$K^{m}$ & Number of UAVs per MNO & 1 \\ \hline
			$|\mathcal{K}^{\text{Tot}}|$ & Total number of UAVs & 5 \\ \hline
			$n$ & Decision threshold for UAV sensing & 3 \cite{zhang2009optimization} \\ \hline
			$B_\circ$ & Total available bandwidth & 100 MHz \\ \hline
			$s$ & Bandwidth per spectrum block & 5 MHz \cite{feng2014flexauc} \\ \hline
			$C$ & Maximum number of available spectrum blocks & 20 \cite{feng2014flexauc} \\ \hline
			$H$ & Flight altitude of UAVs & 200 m \cite{wu2021optimisation} \\ \hline
			$r$ & UAV movement radius & 100 m \cite{wu2021optimisation} \\ \hline
			$\gamma$ & Signal to noise ratio & 18 dB \cite{wu2021optimisation} \\ \hline
			$\lambda$ & Energy detection value & 1.008 \cite{wu2021optimisation} \\ \hline
			\rowcolor{blue!20}
			\multicolumn{3}{|c|}{\textbf{Deep Neural Network}} \\ \hline
			$N_{\text{Episode}}$ & Number of episodes & 2000\\ \hline
			$B$ & Replay buffer size & $10^6$ \\ \hline
			$\mathfrak{B}$ & Batch size & 32 \\ \hline
			$\rho$ & Target network update period & 0.01 \\ \hline
			$H$ & Number of hidden layers & 2 \\ \hline
			$Y_\alpha$ & Number of neurons in each actor hidden layer & 400, 300 \\ \hline
			$Y_\beta$ & Number of neurons in each critic hidden layer & 400, 300 \\ \hline
			$\alpha$ & Actor network learning rate & $10^{-7}$ \\ \hline
			$\beta$ & Critic network learning rate & $10^{-7}$ \\ \hline
			$\gamma$ & Discount factor & 0.9 \\ \hline
		\end{tabular}
	}
\end{table}
\subsection{Numerical Results}
\emph{\textbf{1) Comparison of fairness through deployment of the traditional greedy algorithm and the proposed MSWGA algorithm:}}
\begin{figure}[b]
	\centering 
	\includegraphics[width=0.8\linewidth]{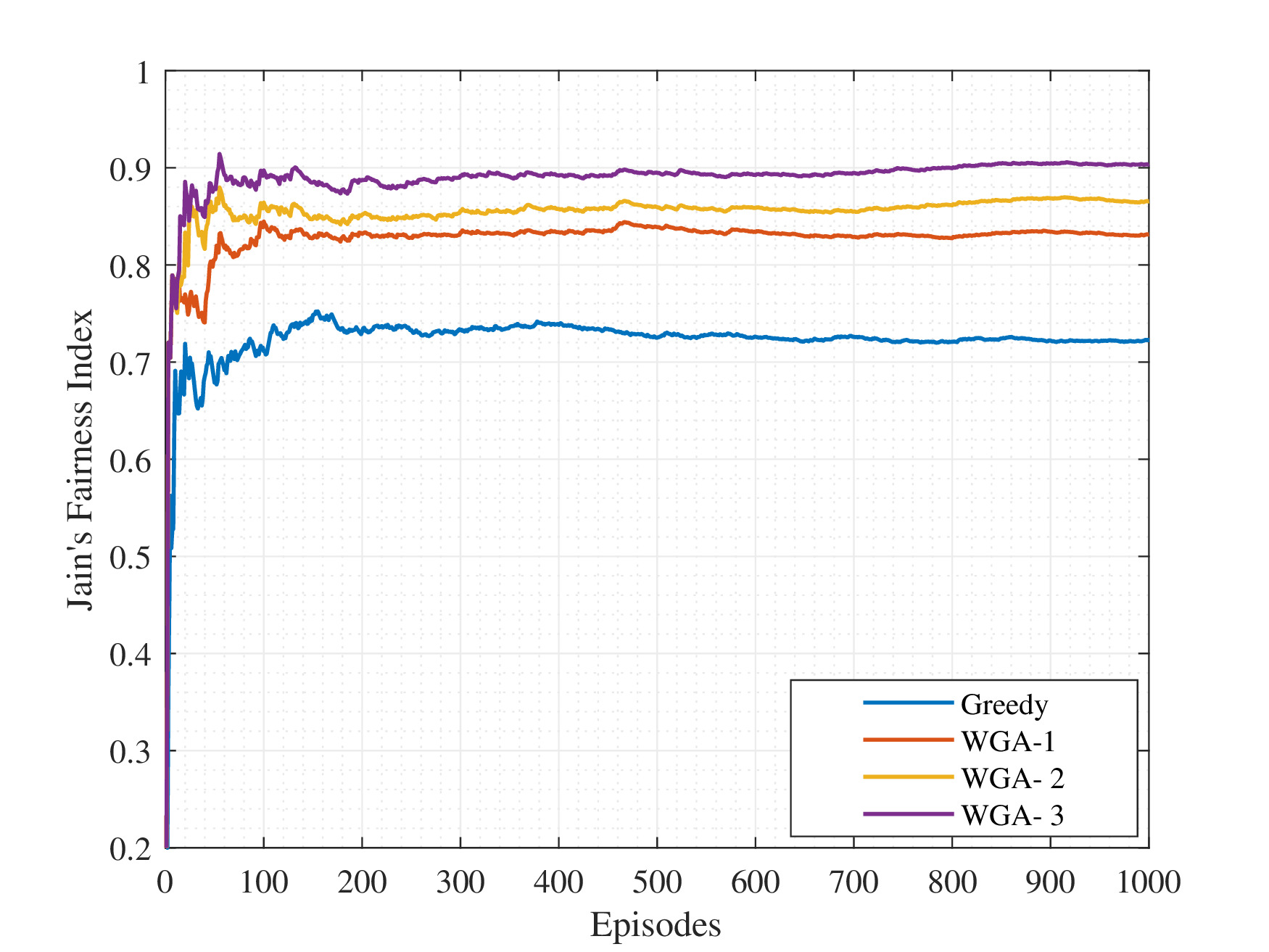}
	\caption{\color{black}{Fairness Index Calculation: Greedy Algorithm and Proportional Weights in Non-RL Scenario.}}
	\label{without learning}
\end{figure} 
We first determine the winners using 
traditional knapsack algorithm. As mentioned before, this algorithm systematically evaluates various combinations to determine the optimal selection for winning MNOs based on the given bids and weights. The resulting fairness index is demonstrated by the blue curve in Fig. \ref{without learning}.
Then, we explore the fairness index in the weighted modes, where different weights were assigned to the bids in each auction. The results, are represented by the red, yellow, and purple curves and correspond to the relationships  (\ref{Weight Based on Wins and Requests}), (\ref{Weight Based on utility}), and (\ref{Combined weight}), respectively.
As can be seen in the figure, the best result happens when considering the ratio of winning times to the number of requests and utility function for the previous auction $(j-1)$ as in formula \eqref{Combined weight}, and iteratively updating these values as the weight of the bids. In this case, \textcolor{black}{
the improvement of the fairness index across the entire network is approximately $27\%$ compared to the traditional greedy method}. This improvement can be attributed to the defined weights, which aim to maintain network fairness over time.

\begin{figure}[]
	\subfigure[\label{market share}]{\includegraphics[width=0.35\linewidth]{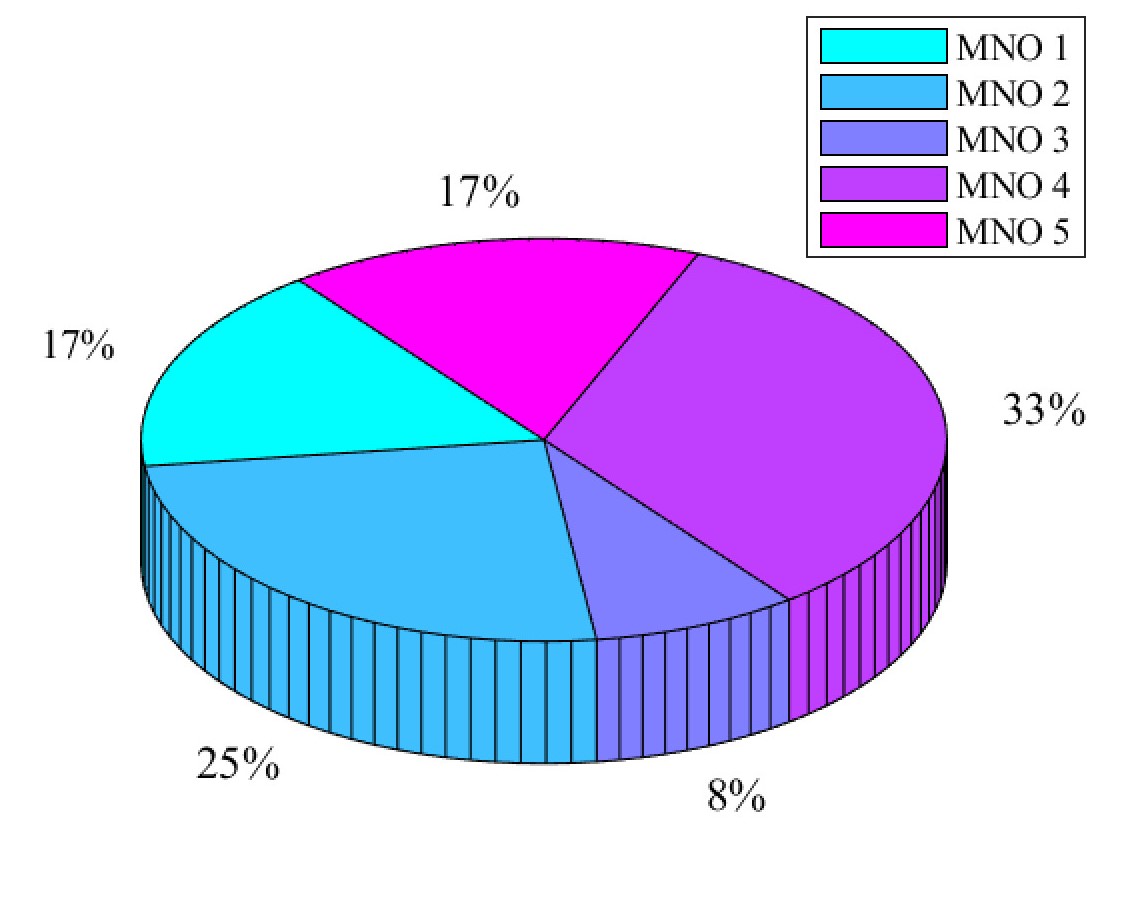}} \,\,
	\subfigure[\label{win_to_requset}]{\includegraphics[width=0.72\linewidth]{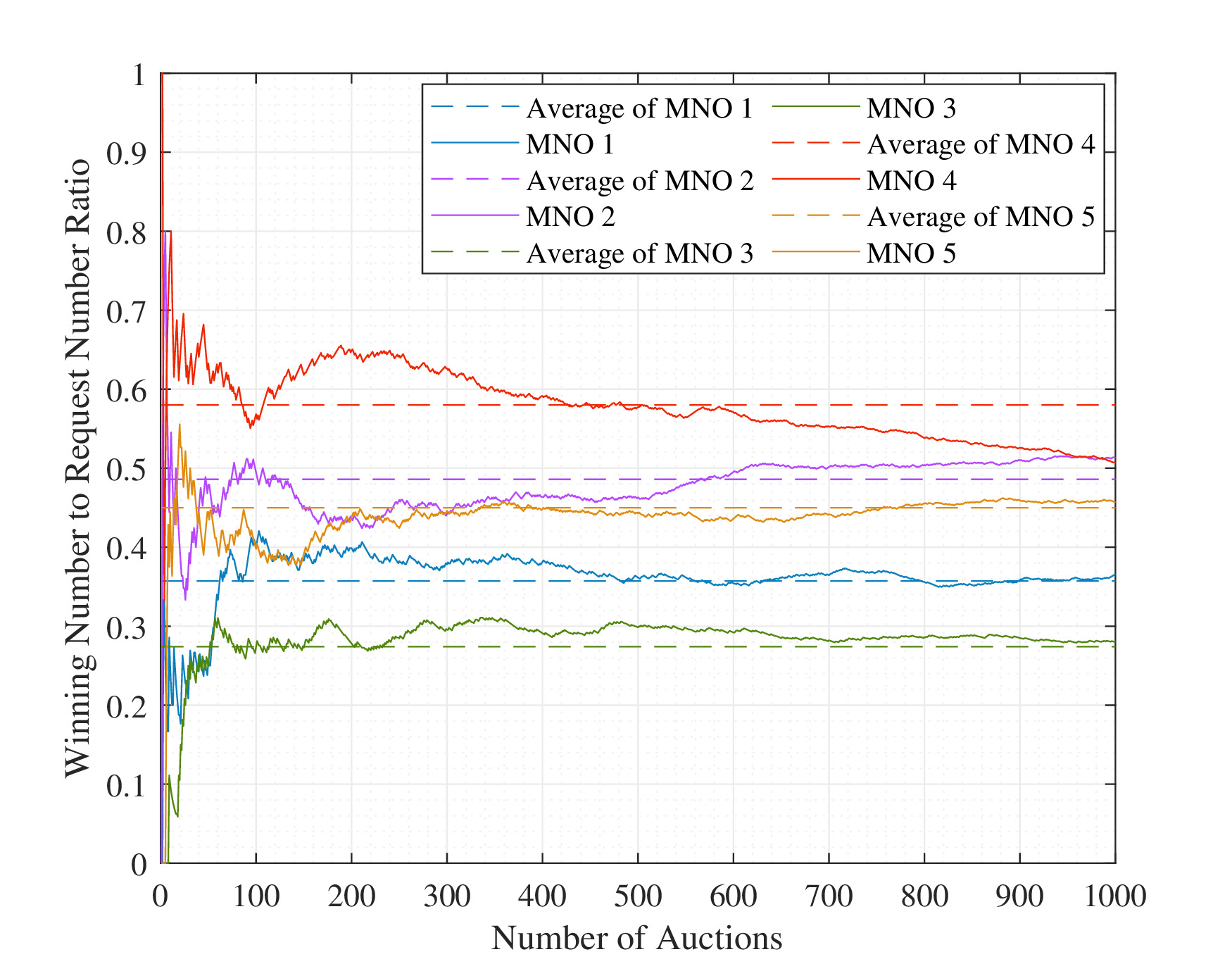}} \,\,
	\caption{\textcolor{black}{Market share of the MNOs \ref{market share}, Effect of market share on win to request ratio of MNOs \ref{win_to_requset}, The number of wins per number of requests ratio for each MNO per auctions.}}
	\label{fig:DATA}
\end{figure}

\emph{\textbf{2) Effect of market share on average fairness per MNO:}} \label{primary coefficient} 
We assume that each operator's market share varies, indicating that fairness in this network does not mean equality of the winning numbers of MNOs. Instead, it involves a combination of their market share and their participation in the previous auctions. To address this, we incorporated the operators' market share as a constant factor in the assigned weights. To ensure meaningful comparisons among the results, we maintained this market share throughout all simulation stages and algorithms, as depicted in Fig. \ref{market share}.
Fig. \ref{win_to_requset} illustrates the individual convergence of each MNO, establishing a direct relationship between the ratio of wins to requests and the respective market share, as demonstrated in Fig. \ref{market share}. To present this correlation effectively, we have obtained the corresponding average values and plotted them with dashed lines. 
By comparing these average values and the market share, it can be inferred that the fairness to all MNOs is adequately provided.
\begin{figure}[b]
	\centering 
	\includegraphics[width=0.8\linewidth]{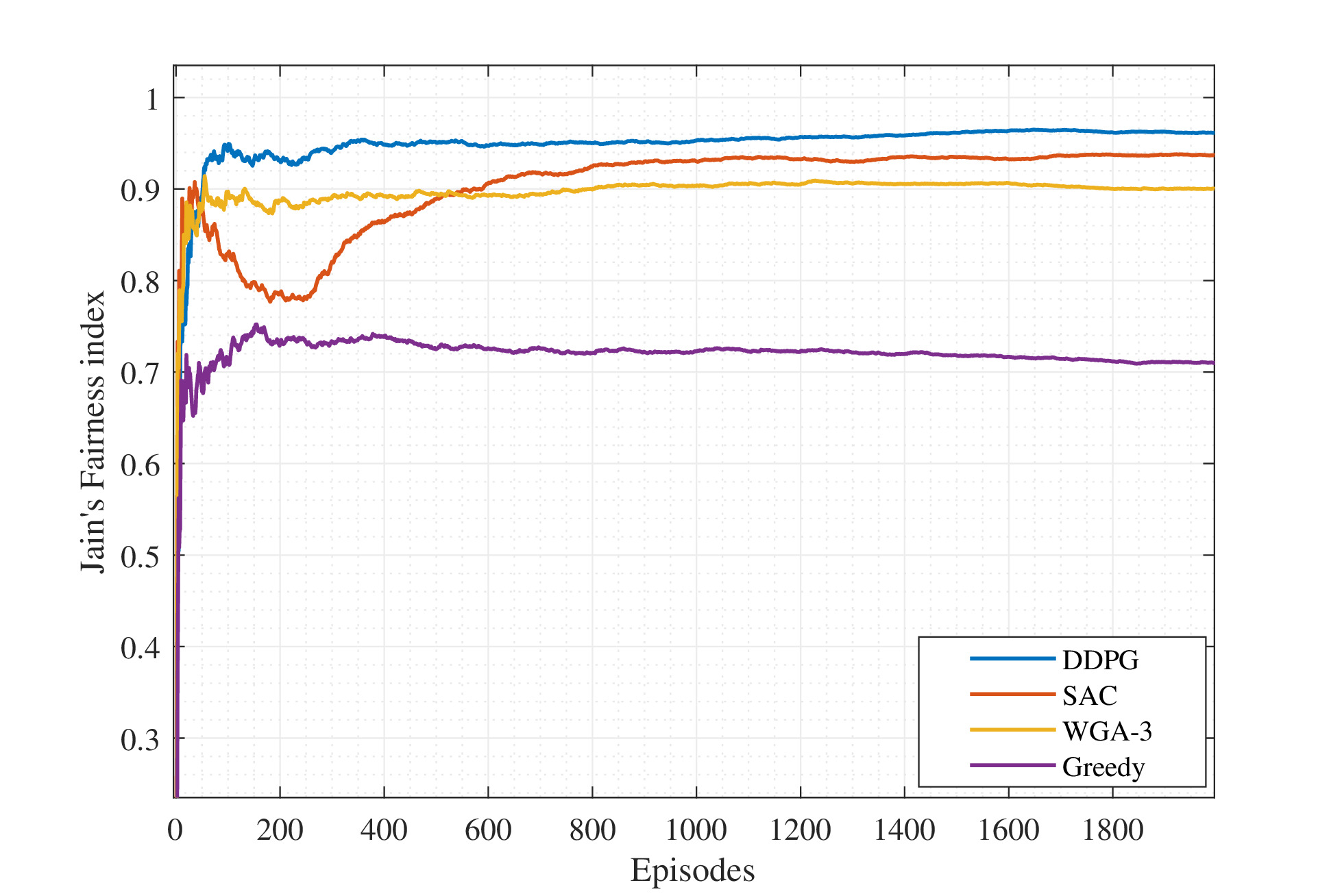}
	\caption{\color{black}{Comparison between learning-based algorithms, greedy algorithm and proposed algorithm.}}
	\label{learning vs greedy}
\end{figure} 

\emph{\textbf{3) Performance comparison between learning-based algorithms and the greedy methods:}}
In this part, we conducted a simulation of the optimization problem using RL methods, i.e., DDPG and SAC to optmizie fairness index. These methods are well-suited for problems with a continuous action space, which align with the research problem at hand. To get the results, the simulation was performed over $2000$ episodes, and the outcomes were compared to the traditional greedy algorithm and the best outcome obtained from the weighted method (WGA-3) obtained from \eqref{Combined weight}.
As depicted in Fig. \ref{learning vs greedy}, the results obtained from the RL approach, particularly using the DDPG algorithm, outperforms other algorithms. Overall, compared to traditional greedy method the use of RL yielded approximately a $\%35$ improvement in the fairness index of the network. However, when compared to WGA-3 method, this improvement is limited to $\%8$.
\textcolor{black}{{It is important to note that such enhancements in fairness index are achieved while maintaining the network performance.} }
 
\begin{figure}[t]
	\centering 
	\includegraphics[width=0.8\linewidth]{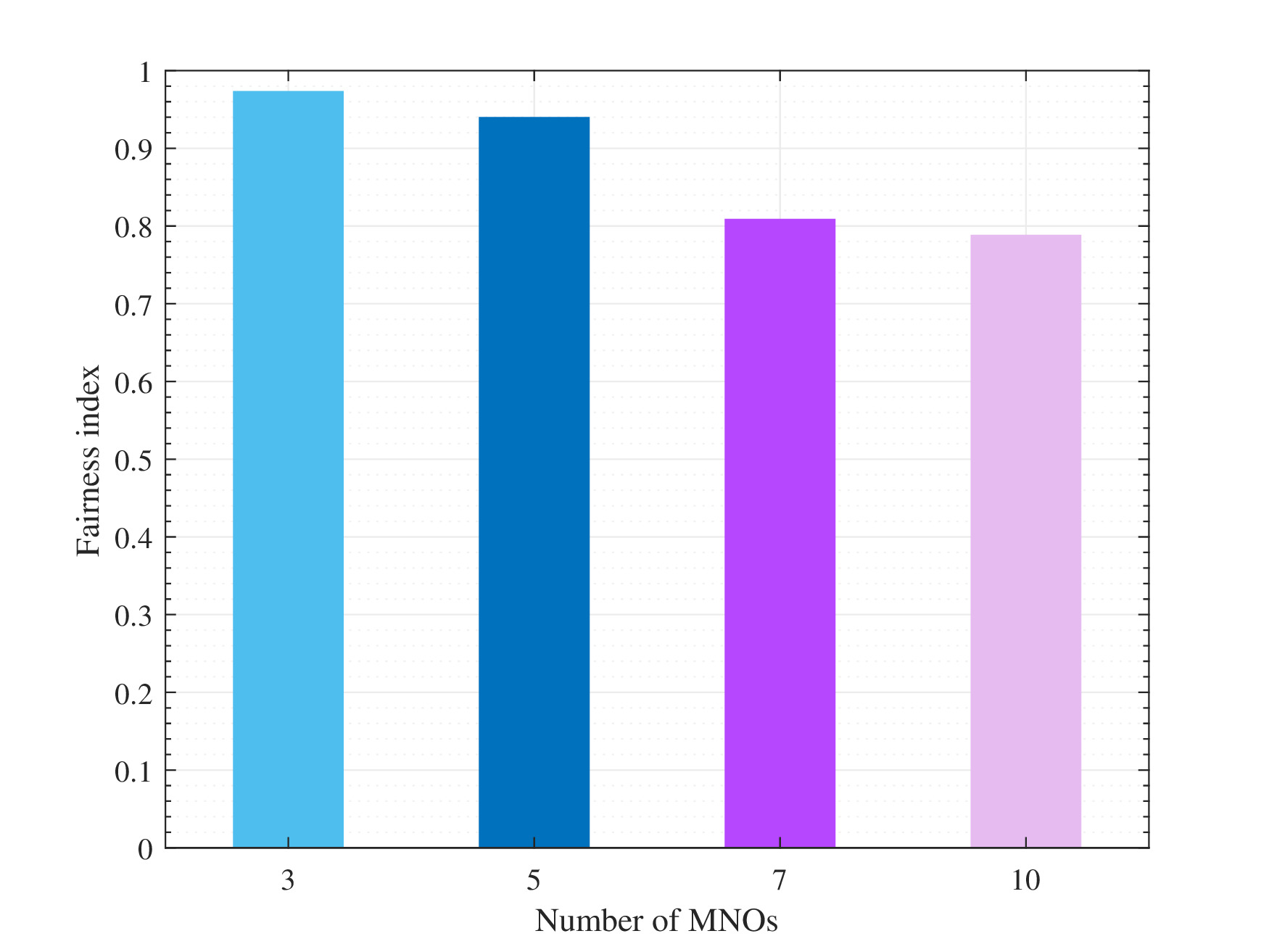}
	\caption{\color{black}{The effect of various MNO numbers on fairness index of the network.}}
	\label{MNO_scenarios}
\end{figure}

 \emph{\textbf{4) Effect of the network size on fairness:}}
An important aspect of the greedy algorithm is that the fairness does not get significantly affected by the network size, i.e., the number of MNOs. However, this is not necessarily the case for the DDPG algorithm. As such, we have to evaluate the effect of network size on fairness index in this case. As shown in Fig \ref{MNO_scenarios}, the number of MNOs does have a meaningful impact on the fairness index, i.e., it reduces it from almost prefect fairness for 3 MNOs to below $80\%$ for 10 MNOs. Nevertheless, please note that the number of MNOs are usually much lower than 10. For smaller number of MNOs, the fairness index is between $80\%$ and $90\%$, which is comparable to that of the greedy algorithm (see Fig \ref{without learning}).

\begin{figure}[t]
	\centering 
	\includegraphics[width=0.8\linewidth]{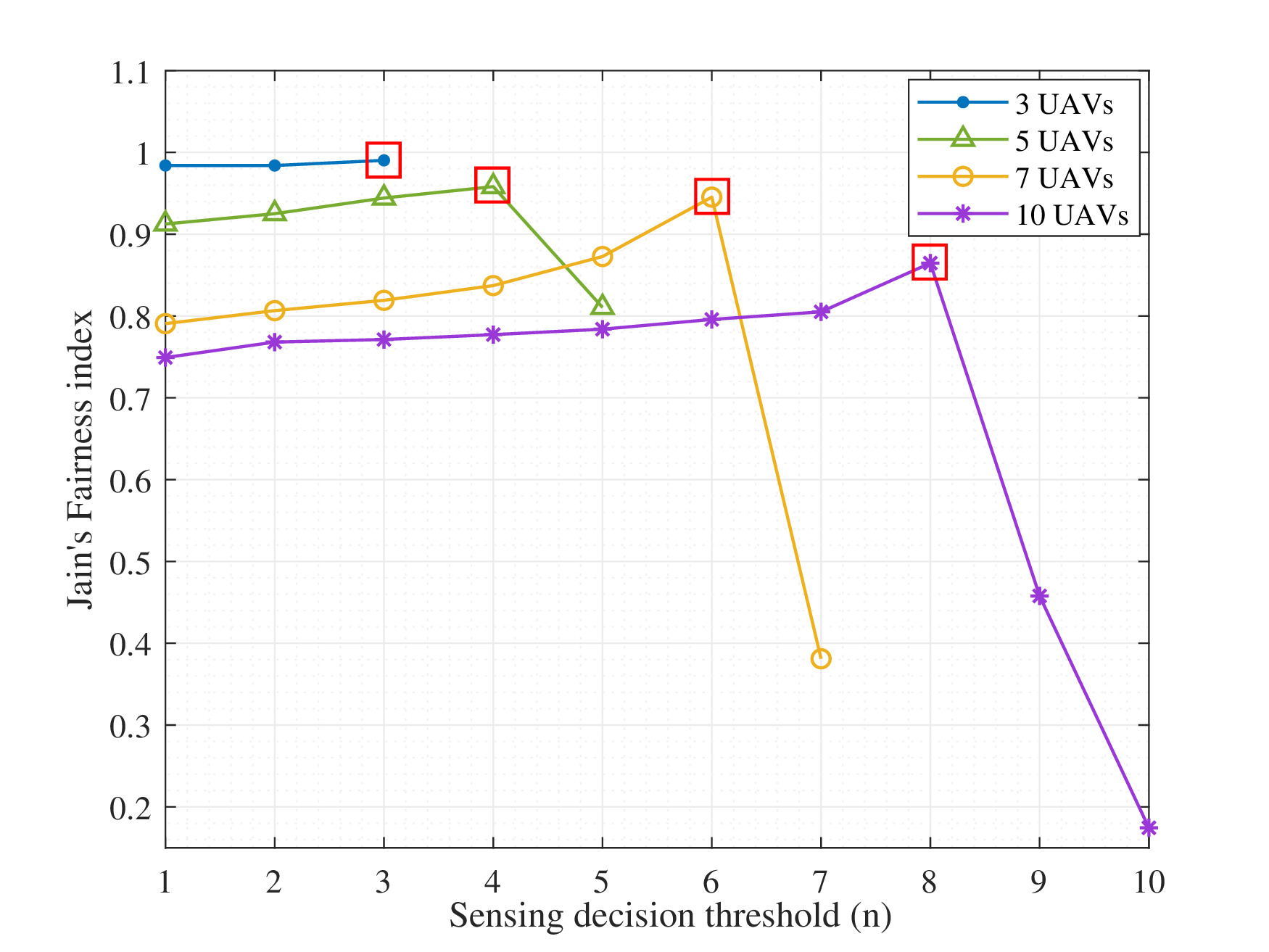}
	\caption{\color{black}{The effect of various sensing decision threshold (n) on the network fairness index.}}
	\label{n_sense}
\end{figure}

\indent \emph{\textbf{5) Effects of sensing decision threshold:}}
In this part of the simulation, we investigate the impact of sensing decision threshold on the fairness index within the ELSA network.
To assess this effect, we conducted simulations for different network configurations, involving $3$, $5$, $7$ and $10$ UAVs (MNOs). For each configuration, we varied the parameter `$n$'. Fig. \ref{n_sense} illustrates the results. Notably, a higher sensing decision threshold can lead to a significant reduction in the fairness index within the network.
The reason behind this outcome lies in the online auction process, where the LBM expects  a certain number of UAVs or all of them. When a larger number of UAVs report identical results to the LBM, the number of wins decreases, and due to the nature of this online auction, the number of requests increases. Consequently, the fairness index, which is defined based on the ratio of wins to requests, experiences a substantial decrease.
It appears that an appropriate selection for the parameter `$n$' follows the relation $n < 	|\mathcal{K}^{\text{Tot}}|$, but the value of $n$ should lean towards $|\mathcal{K}^{\text{Tot}}|$, e.g., $|\mathcal{K}^{\text{Tot}}|-1$ or $|\mathcal{K}^{\text{Tot}}|-2$.
\section{Conclusion}\label{VI}
In this paper, a weighted-based spectrum auction mechanism (FVCG) by considering the corresponding market share of MNOs was developed for an ELSA system, aiming at maximizing the fairness index of the system. We proposed two methods to achieve the maximum fairness of the system; greedy-based and RL-based algorithms. Both used the history of previous auctions to determine the weight of bids. Furthermore, we defined some different weights for greedy-based algorithms and considered the weight of each bid as the action of RL-based algorithms. 
The results obtained from our proposed approach, demonstrate that we have been able to achieve the target fairness, e.g, for a network with $5$ MNOs, we showed that the fairness index can converge to approximately $\%90$ and $\%98$, for MSWGA and DDPG algorithms,  respectively.
The MSWGA mechanism, based on greedy algorithms, yielded close to $\%27$ enhancement compared to the traditional greedy algorithm used in VCG auctions. This enhancement is increased to $\%35$ if we deploy DRL-based method of DDPG. 
Furthermore, our investigation of the decision threshold's impact on the fairness index using the cooperative sensing method revealed insightful achievements.
In particular, if we are using $|\mathcal{K}^{\text{Tot}}|$ cooperative sensors (e.g. UAV's), choosing the threshold $n$  equal to $|\mathcal{K}^{\text{Tot}}|$ will drastically degrade the fairness index. An optimal value for $n$ is somewhere close to $|\mathcal{K}^{\text{Tot}}|$ but always less than it. Additionally, when considering market share among MNOs, it becomes evident that fairness in this network relies on each MNOs' respective market share. All in all, in this research, we were able to significantly improve 
the fairness index within the ELSA network proposed in ADEL project. This was achieved by letting the agent effectively track the MNOs' past performance, allowing for the selection of more appropriate bid weights.
\bibliographystyle{ieeetr}
\bibliography{Citation}
\end{document}